\pdfoutput=1
\documentclass{aa}

\usepackage{ulem}
\usepackage{amsmath,amssymb}
\usepackage[varg]{txfonts}

\usepackage{natbib}

\usepackage{color}
\usepackage{graphicx}

\usepackage{epstopdf}

\usepackage{tablefootnote}

\usepackage{gensymb}
\usepackage{stfloats}

\usepackage[utf8]{inputenc}
\usepackage{hyperref}
\usepackage{booktabs,multirow}

\definecolor{mygray}{gray}{0.6}
\definecolor{darkblue}{rgb}{0.0, 0.0, 0.8}

\newcommand{\se}[1]{Sect.~\ref{sec:#1}}
\newcommand{\eq}[1]{Eq.~(\ref{eq:#1})}
\newcommand{\eqs}[2]{Eqs.~(\ref{eq:#1}) and (\ref{eq:#2})}
\newcommand{\Eq}[1]{Equation~(\ref{eq:#1})}
\newcommand{\fg}[1]{Fig.~\ref{fig:#1}}
\newcommand{\Fg}[1]{Figure~\ref{fig:#1}}
\newcommand{\tb}[1]{Table~\ref{tab:#1}}
\newcommand{\Tb}[1]{Table~\ref{tab:#1}}

\begin{document}

\title{Planetesimal formation near the snowline: in or out?}
\author{Djoeke Schoonenberg and Chris W.~Ormel}
\institute{Anton Pannekoek Institute for Astronomy, University of Amsterdam, Science Park 904, 1090 GE Amsterdam, The Netherlands\label{inst1} \\
\email{[d.schoonenberg@uva.nl,c.w.ormel@uva.nl]}}
\date{Received 4 November 2016 \slash~Accepted 4 February 2017}

\abstract{The formation of planetesimals in protoplanetary disks is not well-understood. Streaming instability is a promising mechanism to directly form planetesimals from pebble-sized particles, provided a high enough solids-to-gas ratio. However, local enhancements of the solids-to-gas ratio are difficult to realize in a smooth disk, which motivates the consideration of special disk locations such as the snowline -- the radial distance from the star beyond which water can condense into solid ice.

In this article we investigate the viability of planetesimal formation by streaming instability near the snowline due to water diffusion and condensation. We aim to identify under what disk conditions streaming instability can be triggered near the snowline.

To this end, we adopt a viscous disk model, and numerically solve the transport equations for vapor and solids on a cylindrical, 1D grid. We take into account radial drift of solids, gas accretion on to the central star, and turbulent diffusion. We study the importance of the back-reaction of solids on the gas and of the radial variation of the mean molecular weight of the gas. Different designs for the structure of pebbles are investigated, varying in the number and size of silicate grains. We also introduce a semi-analytical model that we employ to obtain results for different disk model parameters.

We find that water diffusion and condensation can locally enhance the ice surface density by a factor 3--5 outside the snowline. Assuming that icy pebbles contain many micron-sized silicate grains that are released during evaporation, the enhancement is increased by another factor $\sim$$2$. In this `many-seeds' model, the solids-to-gas ratio interior to the snowline is enhanced as well, but not as much as just outside the snowline.
In the context of a viscous disk, the diffusion-condensation mechanism is most effective for high values of the turbulence parameter $\alpha$ ($10^{-3}$--$10^{-2}$). Therefore, assuming young disks are more vigorously turbulent than older disks, planetesimals near the snowline can form in an early stage of the disk. In highly turbulent disks, tens of Earth masses can be stored in an annulus outside the snowline, which can be identified with recent ALMA observations.

}


\maketitle

\section{Introduction}
Planets form in gaseous disks around young stars. Initially, micron-sized dust grains are present in such a disk. It is generally thought that planets form from planetesimals of the order of a kilometer in size, which are large enough for gravity to play an important role \citep{Safronov1969,PollackEtal1996,2000SSRv...92..279B}. However, the first step in the planet formation process, the coagulation of dust grains into planetesimals, is still an unsolved problem. Although micron-sized grains can quite easily coagulate to mm-sized particles \citep{1997ApJ...480..647D,2007A&A...461..215O}, typical relative velocities quickly become so large that particles fragment or bounce rather than stick \citep{2000Icar..143..138B,2010A&A...513A..57Z}. The typical size at which coagulation is no longer feasible is referred to as the fragmentation or bouncing barrier. Observations at radio wavelengths indeed infer a large reservoir of these mm-to-cm size particles (pebbles) \citep[\textit{e.g.}][]{TestiEtal2014}.

Even if it were possible to coagulate uninterruptedly from micron-sizes to large bodies, the growth process would face another challenge: that of radial drift. As soon as particles reach sizes at which they aerodynamically start to decouple from the gas, they lose angular momentum and spiral towards the star \citep{Whipple1972,Weidenschilling1977}. Radial drift is a fast process and hence poses a challenge to planetesimal formation theories, as planetesimals should form before the solids are lost to the star.

A promising mechanism to quickly form planetesimals directly from pebbles is streaming instability \citep{2005ApJ...620..459Y,2007Natur.448.1022J,2007ApJ...662..627J}. Streaming instability leads to clumping of pebbles that subsequently become gravitationally unstable and form planetesimals. 
In order for streaming instabilities to occur, the metallicity of the disk has to be locally enhanced above the typical value of $1\%$ \citep{2009ApJ...704L..75J,BaiStone2010i,DrazkowskaDullemond2014}. However, if one considers a smooth disk ({\it i.e.}, without any radial pressure bumps or special locations), solids are flushed to the star in an inside-out manner due to dust growth and radial drift (\textit{e.g.}, \citet{BirnstielEtal2010i,2012A&A...539A.148B,2016A&A...586A..20K,SatoEtal2016})\footnote{\citet{2016A&A...594A.105D} argue that growth and radial drift can lead to pile-ups in the inner disk, where streaming instability may then be triggered.}. Radial drift lowers the metallicity of the disk and -- at first sight -- works against creating conditions favorable to streaming instability.

A special location that can affect the distribution of solids in the disk is the snowline. The snowline is the radius from the central star interior to which all water is in the form of vapor. Recent observations of structures in protoplanetary disks (for example HL Tau; \citealt{ALMA-Partnership2015}) have been proposed to be associated with icelines (not necessarily water), \textit{e.g.}, \citet{2015ApJ...815L..15B,2016ApJ...821...82O,2016ApJ...819L...7N}.

The viability of a pile-up of solids at the snowline has been a topic of interest for considerable time. Several mechanisms have been proposed, such as the creation of a pressure maximum \citep[\textit{e.g.}][]{2008A&A...487L...1B} or a deadzone \citep[\textit{e.g.}][]{2007ApJ...664L..55K}. Another pile-up mechanism is water diffusion and condensation: water vapor in the inner disk can diffuse outward across the snowline and condense, thereby increasing the solids reservoir just outside the snowline.  Works that have investigated the role of water diffusion and condensation on the growth and distribution of solids include \citet{1988Icar...75..146S,2004ApJ...614..490C, 2006Icar..181..178C,2013A&A...552A.137R,EstradaEtal2016,ArmitageEtal2016} and, recently, \citet{2016arXiv161006463K}, which focuses on the atmospheric snowline. In this paper, we investigate the role of the radial snowline to enhance the solids-to-gas ratio.

The works most relevant to our problem are \citet{1988Icar...75..146S} and \citet{ 2013A&A...552A.137R}. 
\citet{1988Icar...75..146S} investigated the role of outward diffusion of water vapor across the snowline. In their work, a `cold-finger' effect was considered: water vapor from the inner disk condenses on (pre-existing) solids beyond the snowline.  In this way, they found a steep increase in the solid density beyond the snowline and hypothesized that this could accelerate the formation of Jupiter. However, dynamical aspects of the disk, such as radial drift of solids and accretion of gas, were not taken into account; except for the vapor, their model is essentially static.
More recently, \citet{2013A&A...552A.137R} also simulated pebble growth by condensation, adopting a Monte Carlo method. In their simulation, the purely icy particles grow by condensation, until the point that they are `lost' by evaporation (modelled as an instantaneous process). Because their work does not feature silicate `seeds' that are immune to evaporation, the remaining ice particles can only grow in size. Consequently, a steady-state is never reached, also not because the simulation domain does not connect to the rest of the disk: there is no inflow of small pebbles or removal of vapor-rich gas.

In this work we consider a model that emphasizes the dynamic nature of the disk, characterized by the accretion rates of pebbles and gas. Our model of the iceline therefore includes inflow of pebbles (at the outer boundary) and outflow of vapor (at the inner boundary) -- a design that results in a steady-state.  Our goal is to investigate whether-or-not this more dynamic design will result in a solids enhancement that could trigger streaming instabilities. In particular, we will investigate how the solids-to-gas ratio will depend on the properties of the disk: the accretion rates (of solids and gas), the turbulence strength, and the aerodynamical properties of the drifting pebbles. We will also investigate which side of the snowline the solids-to-gas ratio is boosted most: interior or exterior.

We use a characteristic particle method, assuming a single pebble size at each time and location in the disk \citep{Ormel2014,SatoEtal2016}. We numerically solve the partial differential equations that govern the transport and condensation/evaporation processes for solids and vapor. We consider two model designs for the composition of pebbles: `single seed' (all silicates in one core) and `many seeds' (many tiny silicates grains). Our model also accounts for feedback of solids on the gas (a.k.a.\ collective effects) and feedback of water vapor on the scale height of the disk. A semi-analytic model is presented that approximates the numerical results, enabling us to carry out parameter searches.

We start out with a description of our model in \se{model}. In \se{results-fid} we discuss numerical results for a fiducial set of model parameters in detail. We first describe the time-dependent numerical results in \se{timedependent}, after which we focus on the steady-state solution in \se{results-fid-steadystate}. In \se{results-parstudy} we summarize the semi-analytical approximate model (which is described in more detail in Appendix~C) and present our results concerning streaming instability conditions. We discuss our results in \se{discussion} and present our main conclusions in \se{conclusions}.
A list of frequently used symbols can be found in Table \ref{tab:symbols}.

\section{Model}\label{sec:model}

\subsection{Disk model}
\label{sec:disk-model}
Throughout this paper, we use the $\alpha$-accretion model \citep{1973A&A....24..337S} to model disks. We fix the mass of the central star $M_{\star} = M_{\odot}$. We fix the temperature profile $T (r)$ to:
\begin{equation}\label{eq:temp}
T = 150 \left(\frac{r}{3~\rm{au}}\right)^{-1/2} \: \rm{K}
\end{equation}
where $r$ is the radial distance from the star. This temperature power-law profile corresponds to a passively-irradiated disk and a solar-mass star \citep{1987ApJ...323..714K,ArmitageEtal2016}; for simplicity, we neglect viscous heating. The isothermal sound speed $c_s$ is related to the temperature as:
\begin{equation}
c_s = \sqrt{\frac{k_B T}{\mu}}
\end{equation}
with $k_B$ the Boltzmann constant and $\mu$ the molecular weight of the gas, which is 2.34 times the proton mass for a typical solar metallicity gas. 
The disk scale height $H_{\rm{gas}}$ is given by:
\begin{equation}
H_{\rm{gas}} = \frac{c_s}{\Omega} \approx 0.033 \left(\frac{r}{1~\rm{au}}\right)^{5/4} \: \rm{au}
\end{equation}
where $\Omega$ is the Keplerian orbital frequency. The gas accretes to the central star at a rate $\dot{M}_{\rm{gas}}$. Given $\alpha$ and $\dot{M}_{\rm{gas}}$, the steady-state gas surface density is \citep{1974MNRAS.168..603L}:
\begin{equation}\label{eq:siggas}
\Sigma_{\rm{gas}} = \frac{\dot{M}_{\rm{gas}}}{2 \pi r v_{\rm{gas}}} = \frac{\dot{M}_{\rm{gas}}}{3 \pi \nu}
\end{equation}
\noindent
where $v_{\rm{gas}} = 3 \nu / 2 r$ is the radial speed of the accreted gas, with $\nu$ the turbulent viscosity, which is given by:
\begin{equation}\label{eq:nu}
\nu = \alpha c_{s} H_{\rm{gas}}
\end{equation}
where $\alpha$ is the dimensionless turbulence parameter. As a result of our choices for the temperature and scale height profiles, $\Sigma_{\rm{gas}} (r)$ is a power-law with index $-1$. We take the turbulent gas diffusivity $D_{\rm{gas}}$ equal to the turbulent viscosity $\nu$.

\subsection{Radial motion of ice, silicates and vapor}
In this study, we are interested in the pebble-sized fraction of the solids -- we do not consider the smaller-sized dust. Pebbles are characterized by the fact that they are aerodynamically partly decoupled from the gas disk.
The gas disk is partly pressure-supported, and therefore rotates at a sub-Keplerian velocity. The pebbles do not feel the pressure gradient, and therefore tend to move toward Keplerian orbital velocities. This means that pebbles are moving through a more slowly rotating gas disk, thereby losing angular momentum through gas drag. The loss of angular momentum results in pebbles spiralling (`drifting') towards the central star. The extent to which pebbles drift depends on the stopping time $t_\mathrm{stop}$, which measures how strongly the pebbles are coupled to the gas -- it is the time after which any initial momentum is lost due to gas friction.

Concerning the stopping times of pebbles we can distinguish between two regimes: in the Stokes regime, the particle size is larger than the mean-free path of the gas molecules $l_{\rm{mfp}}$, and the stopping time is calculated in a fluid description; in the Epstein regime, the particle size is smaller than the mean-free path of the gas molecules, and a particle description is needed instead. The mean-free path $l_{\rm{mfp}}$ is given by:
\begin{equation}\label{eq:lmfp}
l_{\rm{mfp}} = \frac{\mu}{\sqrt{2} \rho_{\rm{gas}} \sigma_{\rm{mol}}}
\end{equation}
where $\sigma_{\rm{mol}}$ is the molecular collision cross-section, and $\rho_{\rm{gas}}$ is the gas density. We take $\sigma_{\rm{mol}} = 2 \times 10^{-15} \: \rm{cm}^2$
as the collisional cross-section of molecular hydrogen \citep{chapman1970mathematical}. 
The stopping time is given by:
\begin{equation}\label{eq:stoppingtime}
t_\mathrm{stop} = \begin{cases} \displaystyle \frac{\rho_{\bullet,\rm{p}} s_{\rm{p}}}{v_{\rm{th}} \rho_{\rm{gas}}}, &\text{(Epstein:}\: \: \: s_{\rm{p}} < \frac{9}{4} l_{\rm{mfp}})\\[1em]
\displaystyle \frac{4 \rho_{\bullet,\rm{p}} s_{\rm{p}}^{2}}{9 v_{\rm{th}} \rho_{\rm{gas}}}, &\text{(Stokes:}\: \: \: s_{\rm{p}} > \frac{9}{4} l_{\rm{mfp}})
\end{cases}
\end{equation}
where $s_{\rm{p}}$ is the particle radius, $\rho_{\bullet,\rm{p}}$ is the particle internal density, and $v_{\rm{th}}$ is the thermal velocity of the gas molecules, defined as $v_{\rm{th}} = \sqrt{8/\pi} c_{s}$. The dimensionless stopping time (sometimes referred to as the Stokes number) is $\tau_{S} = t_\mathrm{stop}\Omega$.

The radial drift velocity of pebbles $v_{\rm{peb}}$ is given by \citep{Weidenschilling1977,NakagawaEtal1986}:
\begin{equation}\label{eq:vdriftnoce}
v_{\rm{peb}} = \frac{v_{\rm{gas}} + 2 \eta v_K \tau_{S}}{1 + \tau_{S}^{2}}
\end{equation}
where $\eta v_K$ is the magnitude of the azimuthal motion of the gas disk below the Keplerian velocity $v_K$:
\begin{equation}
    \label{eq:eta}
\eta v_{K} = -\frac{1}{2} \frac{c_{s}^{2}}{v_{K}} \frac{\partial \log P}{\partial \log r}
\end{equation} 
with $P$ the gas pressure.
Under normal disk conditions, the solids-to-gas ratio is about 1 in 100, and it is reasonable to neglect the back-reaction of the solids on the gas. In this paper we are interested in situations where the solids-to-gas ratio is boosted, however, and therefore we study the effects of the back-reaction of solids on the gas, or `collective effects', as well. Collective effects have been calculated by \citet{NakagawaEtal1986}. However, their formula was derived in the inviscid limit. In Appendix~\ref{app:collective} we have calculated the drift velocity, accounting for both the backreaction and viscous forces. When including the back-reaction, we replace \eq{vdriftnoce} by \eq{coll-eqs-final}.

Water vapor (denoted by the subscript $Z$) mixes with the gas and gets accreted to the central star at the same velocity as the gas.

\subsection{Evaporation and condensation}
The rate of change in mass ($dm_{\rm{p}}/dt$) of a spherical icy pebble is given by \citep{1991Icar...90..319L,2006Icar..181..178C,2013A&A...552A.137R}:
\begin{equation}\label{eq:dmdt}
\frac{d m_{\rm{p}}}{d t} = 4 \pi s_{\rm{p}}^2 v_{\rm{th}, Z} \: \rho_{Z} \left(1 - \frac{P_{\rm{eq}}}{P_{Z}}\right)
\end{equation}
where $s_{\rm{p}}$ is the radius of the pebble, $v_{\rm{th}, Z}$ is the thermal velocity of vapor particles, $\rho_{Z}$ is the vapor density, $P_{Z}$ is the vapor pressure, and $P_{\rm{eq}}$ is the saturated or equilibrium pressure, which is given by the Clausius-Clapeyron equation:
\begin{equation}\label{eq:clauclap}
P_{\text{eq}} = P_{\rm{eq, 0}} e^{- T_{a} / T}
\end{equation}
where $T_{a}$ and $P_{\rm{eq, 0}}$ are constants depending on the species. For water, $T_{a}$ = 6062 K and $P_{\rm{eq, 0}} = 1.14 \times 10^{13} \: \rm{g} \: \: \rm{cm}^{-1} \: \: \rm{s}^{-2}$ \citep{1991Icar...90..319L}. 

Assuming spherical pebbles and one typical particle mass $m_{\rm{p}}$ at each location and time in the disk (\textit{i.e.}, the particle size distribution is narrow), and using the ideal gas law to write $P_{Z}$ in terms of the (midplane) vapor density $\rho_{Z}$, we can rewrite \eq{dmdt} into a change in the ice surface density $\Sigma_{\rm{ice}}$ due to condensation and evaporation:
\begin{equation}\label{eq:icesourceterms}
\dot\Sigma_\mathrm{ice, C/E} = R_{\rm{c}} \Sigma_{\rm{ice}} \Sigma_{Z} - R_{\rm{e}} \Sigma_{\rm{ice}}
\end{equation}
where the dot denotes the time derivative. In \eq{icesourceterms} the vapor surface density $\Sigma_{Z}$ is related to the midplane vapor density as $\rho_{Z} = \Sigma_{Z} / \sqrt{2 \pi} H_{\rm{gas}}$ and $R_{\rm{c}}$ and $R_{\rm{e}}$ are the condensation rate and evaporation rate, respectively, defined as:
\begin{align}\label{eq:ecrates}
R_c &= 8 \sqrt{\frac{k_B T}{\mu_{Z}}} \frac{s_{\rm{p}}^{2}}{m_{\rm{p}} H_{\rm{gas}}} \\
R_e &= 8 \sqrt{2 \pi} \frac{s_{\rm{p}}^{2}}{m_{\rm{p}}} \sqrt{\frac{\mu_{Z}}{k_B T}} P_{\rm{eq}}
\end{align}

\noindent
with $k_B$ the Boltzmann constant, and $\mu_{Z} = 18 m_{\rm{H}}$ the mean molecular weight of water vapor, where $m_{\rm{H}}$ is the proton mass. Evaporating ice turns into vapor, and condensing vapor turns into ice, and therefore we find for the vapor source terms:
\begin{equation}\label{eq:Zsourceterms}
\dot\Sigma_\mathrm{Z, C/E}  
= - \dot\Sigma_\mathrm{ice, C/E}
= - R_{\rm{c}} \Sigma_{\rm{ice}} \Sigma_{Z} + R_{\rm{e}} \Sigma_{\rm{ice}}
\end{equation} 

\subsection{Internal structure and composition of pebbles}\label{sec:pebblecom}
We assume that in the outer disk, half of the mass in pebbles consists of water ice, and the other half consists of silicates \citep{2003ApJ...591.1220L, 2015Icar..258..418M}. That is, the dust-fraction in pebbles in the outer disk $\zeta$ = 0.5. Adopting $\rho_{\bullet,\rm{ice}} = 1 \: \rm{g} \: \rm{cm}^{-3}$ as the density of a pure ice particle and $\rho_{\bullet,\rm{sil}} = 3 \: \rm{g} \: \rm{cm}^{-3}$ as the density of a pure silicate particle, we find that in the outer disk, the internal density of a pebble is $\rho_{\bullet,\rm{p}}= 1.5 \: \rm{g} \: \rm{cm}^{-3}$. Generally, the internal density of an ice/silicate pebble is given by:
\begin{equation}
\rho_{\bullet,\rm{p}} = \frac{\rho_{\bullet,\rm{ice}} \rho_{\bullet,\rm{sil}} (m_{\rm{sil}} + m_{\rm{ice}})}{m_{\rm{sil}} \rho_{\bullet,\rm{ice}} + m_{\rm{ice}} \rho_{\bullet,\rm{sil}}}
\end{equation}
where $m_{\rm{ice}}$ is the mass in ice within the pebble and $m_{\rm{sil}}$ is the mass in silicate within the pebble.
\Eq{dmdt} is valid for a pure ice particle, while we use it for icy pebbles with (a) silicate core(s). We checked that using $m_{\rm{ice}}$ instead of $m_{\rm{p}}$ in the condensation and evaporation rates does not change the results significantly.

We consider two possibilities for the internal structure of icy pebbles (\fg{sironocartoon}).
\begin{itemize}
    \item {\bf Single-seed model}. In this model we view pebbles in the outer disk as balls of ice with a single core made of silicates. When the ice in such a pebble evaporates, the silicate core is left behind. In this framework, the number of pebbles does not change when crossing the snowline.

\item {\bf Many-seeds model}. In this model we view icy pebbles as lots of small silicate particles that are `glued' together by ice. The small silicate particles are released by evaporation, and free silicates can stick onto icy pebbles. We do not `resolve' the internal structure of the pebbles, and only trace the total amount of silicates `locked' into icy pebbles. Following \citet{2011ApJ...728...20S}, we assume that the released silicate particles are of micron size. Because they are so small, the silicates do not undergo any aerodynamic drift and follow the radial motion of the gas.
\end{itemize}

\begin{figure}[t]
	\centering
		\includegraphics[width=88mm]{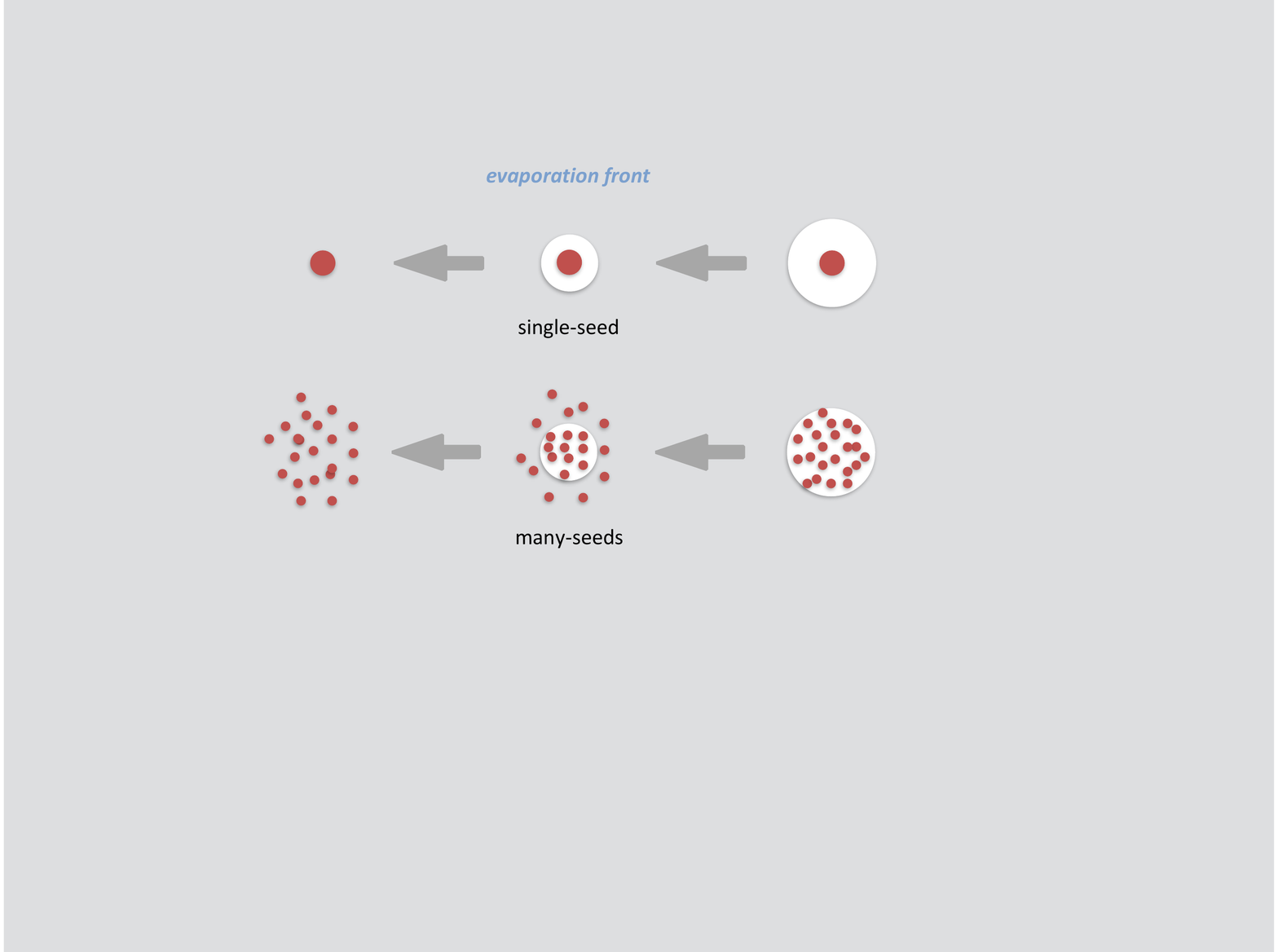}
        \caption{Schematic showing the difference between the single-seed model and the many-seeds model for the internal structure of icy pebbles in the outer disk. In the single-seed model, the water ice layer (white) on the silicate core (red) becomes smaller at the evaporation front, and after complete evaporation of the ice, only one single silicate core remains. Since we take $\zeta = 0.5$, the silicate core contains half the mass of the original pebble: $m_{\rm{core}} = m_{\rm{ice}} = 0.5 m_{\rm{p, start}}$. In the many-seeds model, pebbles in the outer disk consist of many micron-sized silicate grains (red) `glued' together by water ice (white). As the ice evaporates, silicate grains are released as well. After complete evaporation of the ice, many micron-sized silicate grains remain. Again, with $\zeta = 0.5$, the total mass in silicate grains in a pebble in the outer disk is half of the total pebble mass, but this mass is now divided among many silicate grains that each have mass $m_{\rm{sil}}$. See text for more details.\label{fig:sironocartoon}}
\end{figure}

\subsection{Transport equations}
In this subsection we provide the systems of transport equations corresponding to both pebble composition models presented in \se{pebblecom}. Our model is 1D; we integrate over the vertical dimension of the disk and follow column (surface) densities. Midplane densities are calculated assuming a Gaussian distribution in height with scaleheights $H_\mathrm{gas}$ (for the gas and the vapor) and $H_\mathrm{peb}$ (for the pebbles), respectively.

\subsubsection{Single-seed pebble model}
In the single-seed model, we follow the vapor surface density $\Sigma_{Z}$, the ice surface density $\Sigma_{\rm{ice}}$, and the number density of particles $N_{\rm{p}}$. The equations governing the time evolution of these quantities are: 
\begin{align}\label{eq:transport-1seed}
\frac{\partial \Sigma_{\rm{ice}}}{\partial t} &+ \nabla \cdot \mathcal{M}_{\rm{ice}} = \dot{\Sigma}_{\rm{ice, C/E}} \\\label{eq:transport-1seedZ}
\frac{\partial \Sigma_{Z}}{\partial t} &+ \nabla \cdot \mathcal{M}_{Z} = \dot{\Sigma}_{Z, \rm{C/E}} \\\label{{eq:transport-1seedN}}
\frac{\partial N_{\rm{p}}}{\partial t} &+ \nabla \cdot \mathcal{N}_{\rm{p}} = 0 \\\nonumber
\end{align}

\noindent
where the source terms $\dot{\Sigma}_{\rm{ice, C/E}}$ and $\dot{\Sigma}_{Z, \rm{C/E}}$ are defined in \eq{icesourceterms} and \eq{Zsourceterms}, $\mathcal{M}_{\rm{ice}}$ and $\mathcal{M}_{Z}$ are the ice and vapor mass flux, respectively, and $\mathcal{N}_{\rm{p}}$ is the number flux of solid particles. These fluxes are given by:
\begin{align}\label{eq:mass-fluxes}
\mathcal{M}_{\rm{ice}} &= -\Sigma_{\rm{ice}} v_{\rm{peb}} - D_{\rm{p}} \Sigma_{\rm{gas}} \nabla \frac{\Sigma_{\rm{ice}}}{\Sigma_{\rm{gas}}} \\
\mathcal{M}_{Z} &= -\Sigma_{Z} v_{\rm{gas}} - D_{\rm{gas}} \Sigma_{\rm{gas}} \nabla \frac{\Sigma_{Z}}{\Sigma_{\rm{gas}}} \\
\mathcal{N}_{\rm{p}} &= -N_{\rm{p}} v_{\rm{peb}} - D_{\rm{p}} N_{\rm{gas}} \nabla \frac{N_{\rm{p}}}{N_{\rm{gas}}} \\\nonumber
\end{align}

\noindent
where $v_{\rm{peb}}$ is the drift velocity of the pebbles given by \eq{vdriftnoce} or by \eq{coll-eqs-final} when including collective effects, $N_{\rm{gas}}$ is the number of gas particles, and $D_{\rm{p}}$ is the particle diffusivity, which is related to $D_{\rm{gas}}$ through the Schmidt number Sc \citep{2007Icar..192..588Y}: 
\begin{equation}
    D_{\rm{p}} = D_\mathrm{gas} \times \rm{Sc} = \frac{D_\mathrm{gas}}{1 + \tau_{S}^2}
\end{equation}
Since in our models the stopping times are much smaller than unity, we set $D_{\rm{p}}$ equal to $D_{\rm{gas}}$.

Note that a negative flux means that material is transported inwards, towards the star. Under the single-size approximation, the typical pebble mass $m_{\rm{p}}$ is given by:
\begin{equation}
m_{\rm{p}} = m_{\rm core} + \frac{\Sigma_{\rm{ice}}}{N_{\rm{p}}}
\end{equation}
with $m_{\rm{core}}$ the mass of the bare silicate cores of the pebbles.

\subsubsection{Many-seeds pebble model}
We implement the many-seeds model as follows. We now consider two additional populations of solids: dirt and free silicates. Dirt refers to the silicates that are captured in icy pebbles, whereas free silicates are small silicate grains without any ice coating. The transport equations for $\Sigma_{\rm{ice}}$, $\Sigma_{Z}$, and $N_{\rm{p}}$ are the same as in the single-seed pebble model (\eq{transport-1seed}--(19)). The transport equations for the dirt surface density $\Sigma_{\rm{dirt}}$ and free silicates surface density $\Sigma_{\rm{sil}}$ are given by:
\begin{align}\label{eq:transportDIRT}
\frac{\partial \Sigma_{\rm{dirt}}}{\partial t} + \nabla \cdot \mathcal{M}_{\rm{dirt}} &= - \rm{max}\left[ 0, ({\it R}_{\rm{e}} - {\it R}_{\rm{c}} \Sigma_{{\it Z}})\right] \Sigma_{\rm{dirt}} + {\it R}_{\rm{s}} \Sigma_{\rm{sil}}\\
\frac{\partial \Sigma_{\rm{sil}}}{\partial t} + \nabla \cdot \mathcal{M}_{\rm{sil}} &= \rm{max}\left[ 0, ({\it R}_{\rm{e}} - {\it R}_{\rm{c}} \Sigma_{{\it Z}})\right] \Sigma_{\rm{dirt}} - {\it R}_{\rm{s}} \Sigma_{\rm{sil}}\\\nonumber
\end{align}
where $\mathcal{M}_{\rm{dirt}}$ and $\mathcal{M}_{\rm{sil}}$ are the dirt and free silicate mass fluxes, respectively. Free silicates are released from the pebbles at a rate proportional to the decrease of ice surface density due to evaporation and to the amount of dirt that can evaporate from within the pebbles: $(R_{e} - R_{c} \Sigma_{\rm{vap}}) \Sigma_{\rm{dirt}}$, but only when there is net evaporation of ice.
Silicate grains stick to icy pebbles at a collision/sticking rate $R_{\rm{s}}$:
\begin{equation}
R_{\rm{s}} = \frac{\Sigma_{\rm{ice}} + \Sigma_{\rm{dirt}}}{\sqrt{2 \pi} H_{\rm{sil}} m_{\rm{p}}} \cdot \Delta v_{\rm{sil, peb}} \cdot \pi s_{\rm{p}}^{2} 
\end{equation}
where $\Delta v_{\rm{sil, peb}}$ is the relative velocity between the pebbles and the silicates and $H_{\rm{sil}}$ is the scale height of the free silicates. Since the inward drift velocity of the pebbles is much larger than the outward diffusion velocity of the silicates, we take $\Delta v_{\rm{sil, peb}} \approx v_{\rm{peb}}$\footnote{For simplicity, we have ignored other contributions to $\Delta v_{\rm{sil, peb}}$. Turbulent velocities, however, may dominate over radial drift motion when $\alpha$ is large \citep{2003Icar..164..127C,2007A&A...466..413O}. Accounting for turbulent relative velocities will result in an increase of the peak solids-to-gas ratio of $\sim$$10\%$ ({\it e.g.}, $20\%$ for our fiducial parameters presented in Table~\ref{table:fiducial}).}.

The dirt and silicate mass fluxes, $\mathcal{M}_{\rm{dirt}}$ and $\mathcal{M}_{\rm{sil}}$, are given by:
\begin{align}
    \mathcal{M}_{\rm{dirt}} &= -\Sigma_{\rm{dirt}} v_{\rm{peb}} - D_{\rm{p}} \Sigma_{\rm{gas}} \nabla \frac{\Sigma_{\rm{dirt}}}{\Sigma_{\rm{gas}}} \\
    \mathcal{M}_{\rm{sil}} &= -\Sigma_{\rm{sil}} v_{\rm{gas}} - D_\mathrm{gas} \Sigma_{\rm{gas}} \nabla \frac{\Sigma_{\rm{sil}}}{\Sigma_{\rm{gas}}}\label{eq:sirono-massfluxes} 
\end{align}
where $v_{\rm{peb}}$ is again the radial drift velocity of the pebbles, and silicates are advected with the gas.

\subsection{Effects of variable $\mu$}
When icy pebbles evaporate, the gas becomes enriched in water vapor. This means that the mean molecular weight $\mu$ of the gas increases:
\begin{equation}
\frac{1}{\mu} = \frac{f_{\mathrm{H_{2}O}}}{\mu_{\mathrm{H_{2}O}}} +  \frac{f_{\mathrm{gas}}}{\mu_{\mathrm{gas}}}
\end{equation}
where $f_{\mathrm{H_{2}O}}$ is the mass fraction of water molecules, $\mu_{\mathrm{H_{2}O}} = 18 m_{\rm{H}}$ is the mean molecular weight of water, and $f_{\mathrm{gas}}$ is the mass fraction of gas molecules (without the vapor contribution), with $\mu_{\mathrm{gas}} = 2.34 m_{\rm{H}}$.

Since the sound speed $c_s$ is proportional to $\sqrt{1/\mu}$, the sound speed decreases as more water vapor is added. The disk scale height $H_{\rm{gas}}$ is proportional to $c_s$ and will therefore decrease as well: the disk becomes more vertically compact with increasing water vapor. An increase in $\mu$ across the water evaporation front has an effect on the pebble drift velocity $v_{\rm{peb}}$ (\eq{vdriftnoce}) in this region, since $v_{\rm{peb}}$ is proportional to $1 / \mu$ through its dependence on $c_s$, and sensitive to variations in the gas pressure $P$ through $\partial \log P / \partial \log r$. The total gas pressure (hydrogen/helium plus water) is given by:
\begin{equation}
P = P_{\rm{gas}} + P_{Z} = \frac{\Sigma_{\rm{gas}} k_B T}{\sqrt{2\pi} H_{\rm{gas}} \mu_{\mathrm{H_{2} / He}}} + \frac{\Sigma_{Z} k_B T}{\sqrt{2\pi} H_{\rm{gas}} \mu_{\mathrm{H_{2}O}}} 
\end{equation}
We take the viscosity $\nu$ to be independent of variations in $\mu$. Therefore, $\Sigma_{\rm{gas}} = \dot{M}_{\rm{gas}} / 3\pi\nu$ is independent of $\mu$ as well. However, the $\mathrm{H_{2}O}$-enriched gas is characterized by a reduced scaleheight, since $H_{\rm{gas}} \propto \sqrt{1/\mu}$, increasing the midplane gas pressure $P_{\rm{gas}}$ as $P_{\rm{gas}} \propto \sqrt{\mu}$. Evaporation hence increases the pressure gradient and the headwind of the gas through \eq{eta}: pebbles tend to drift faster. Finally, the drift velocity $v_{\rm{peb}}$ also depends on the stopping time $\tau_{S}$, which in turn depends on the gas density and the thermal velocity -- both quantities that depend on $\mu$. We take these dependencies into account as well.

We now define two models that we study in the following sections. In the {\bf simple} model, we do not include collective effects and the variability of the mean molecular weight $\mu$. In the simple model, the sound speed and scale height are independent of variations in $\mu$, $\eta v_K$ (\eq{eta}) is constant and the midplane gas density is $\mu$-independent: $\rho_{\rm{gas}} = \Sigma_{\rm{gas}}/\sqrt{2\pi} H_{\rm{gas}}$.
In the {\bf complete} model, we do take collective effects and variations of $\mu$ into account, as described above.

\subsection{Input parameters}
\label{sec:keypars}
The four key input parameters that can be varied in our model are:

\begin{itemize}
\item The gas accretion rate $\dot{M}_{\rm{gas}}$.
\item The ratio between the pebble accretion rate and the gas accretion rate, denoted by the symbol $\mathcal{F}_{s/g}$, and defined as:
\begin{equation}
\mathcal{F}_{s/g} \equiv \dot{M}_{\rm{peb}} / \dot{M}_{\rm{gas}}.
\end{equation}
\item The dimensionless turbulence parameter $\alpha$.
\item The initial stopping time of pebbles just outside the snowline (at 3~au) denoted by the symbol $\tau_3$. This parameter is simply a proxy for the mass of the incoming pebbles.
\end{itemize}
Together, these input parameters define the `zero-model' of the disk: the gas and solids distribution in the case of no condensation and evaporation. The gas surface density profile $\Sigma_{\rm{gas}}$ is defined through \eq{siggas}, where the turbulent viscosity $\nu$ is determined by $\alpha$ through \eq{nu}. With $\Sigma_{\rm{gas}}$ and the stopping time normalization $\tau_3$ at 3~au, we can find $\tau (r)$ at any other radius $r$. We can then also determine the pebbles surface density $\Sigma_{\rm{peb}}$ for the zero-model: it is given by $\mathcal{F}_{s/g} \dot{M}_{\rm{gas}} / 2\pi r v_{\rm{peb}} (r)$, where the pebble velocity $v_{\rm{peb}} (r)$ depends on $\tau (r)$ and is given by \eq{vdriftnoce} (without backreaction) or \eq{coll-eqs-final} (including backreaction).

\subsection{Model assumptions}\label{sec:assumptions}
In constructing our model we employed several assumptions, which we discuss below.
\begin{itemize}
\item We do not include a particle size distribution, but rather consider one typical pebble size at each time and location in the disk. In the many-seeds model, we do distinguish between the population of small silicate grains and large icy pebbles, however. We come back to the importance of condensation onto small grains rather than onto pebbles in the discussion (\se{discussion}).

\item 
Our model does not account for coagulation among pebbles: the enhancement of the solids-to-gas ratio is solely due to diffusion and condensation. We remain agnostic as to what happens to the pebbles in terms of coagulation and fragmentation until they reach the evaporation front.
In the many-seeds model, we do not take into account coagulation of silicate particles. We will come back to this point in the discussion (\se{discussion}).

    \item Our model is 1D, and has a straight vertical snowline. In reality, the vertical snowline is curved rather than straight, because the temperature varies with height above the midplane ({\it e.g.} \citet{2011ApJ...738..141O}). However, \citet{2013A&A...552A.137R} have shown that particle growth due to diffusion and subsequent condensation of water across the vertical snowline is negligible compared to the same process across the radial snowline.

\item Similarly, we assume that the scale height of the vapor -- and, in the many-seeds implementation, of the free silicates -- is always equal to the scale height of the disk; in other words, when an icy pebble evaporates, the released water vapor immediately distributes itself over the same vertical extent as the background gas, whereas it originates from the icy pebbles that reside in a thinner disk, characterized by a scale height $H_{\rm{peb}}$:
\begin{equation}
H_{\rm{peb}} = H_{\rm{gas}} \sqrt{\frac{\alpha}{\tau_S + \alpha}}.
\end{equation}
The assumption of rapid vertical mixing extends to the calculation of the mean molecular weight of the gas in the complete model. Also there, it is assumed that the water vapor is distributed evenly throughout the gas at all times. In \se{results-fid-steadystate}, we verify that the assumption of taking the vapor scale height equal to the disk scale height rather than the pebbles scale height, does not have a large effect on the results.

    \item We assume a steady-state gas distribution, that is, $\dot M_{\rm{gas}}$ and $T$ remain constant. This assumption is justified since the  solids-to-gas enhancement takes place on a radial scale $\Delta r$ that is small compared to the size of the disk. In other words, the local viscous evolution timescale ($\Delta r^2/\nu$), on which the effect happens, is much smaller than the global viscous evolution timescale of the disk, on which the gas accretion rate and global temperature profile change. The temperature profile (\eq{temp}) corresponds to a passively-irradiated disk. However, in case of viscous heating $T(r)$ will still be described by a power-law \citep{2015arXiv150906382A}. We have checked the effect of adopting a viscous heating temperature profile ($T(r) \propto r^{-3/4}$) in our numerical simulation. The key consequence is that the location of the snowline is shifted in radius; all other model results are very similar.

    \item We adopt an $\alpha$-viscosity model, with constant $\alpha$ (the value of which we allow to vary between very high and very low values), independent of the solids-to-gas ratio. However, if turbulence is driven by the magneto-rotational instability, the local turbulence strength at the midplane depends on the local grain size and abundance. So-called `dead zones', where turbulence is strongly reduced, could occur at locations in the disk with a sharp increase in the solids-to-gas ratio \citep{2007ApJ...664L..55K}. Since our results depend strongly on the amount of turbulence near the snowline, the creation of dead zones would change our results. 
    
    \item Similarly, we do not consider the interplay between (small) grains and the local temperature: we do not solve for the temperature profile self-consistently, but take it as constant, given by \eq{temp}. However, in the many-seeds model, a large amount of small silicate grains is released in the evaporation front. This means that locally the opacity could increase. Conceivably, this could lead to a steeper temperature profile in the region near the snowline and therefore to a more narrow evaporation front. Including these effects requires a radiation transfer model, which is beyond the scope of this work.
  
\item In the many-seeds implementation, we do not include condensation onto the small, free silicate particles. We also do not model the porosity of pebbles. We come back to these issues in \se{disc-condens}. 
\end{itemize}

\begin{figure*}[t]
	\centering
		\includegraphics[width=0.495\textwidth]{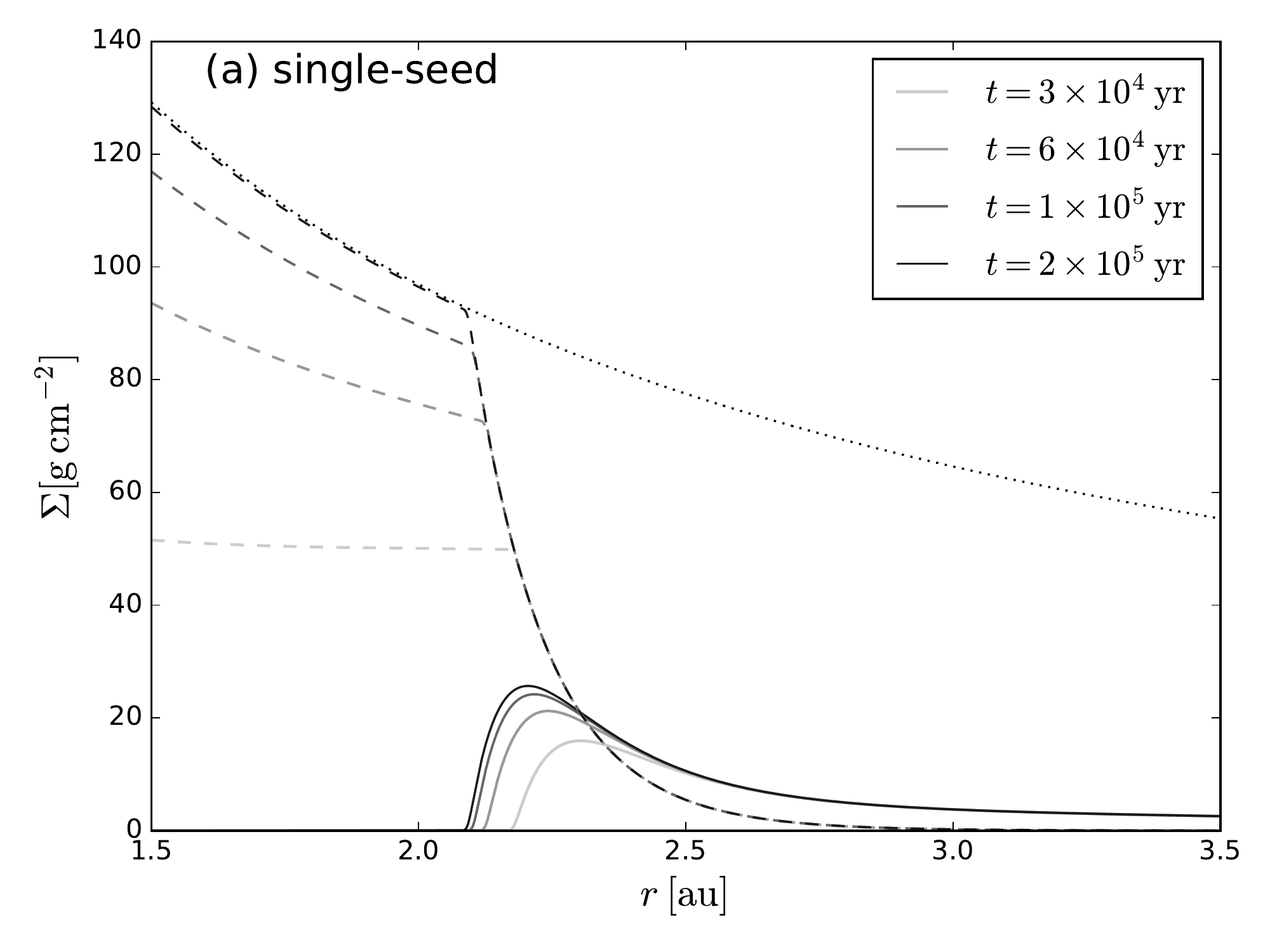}
		\includegraphics[width=0.495\textwidth]{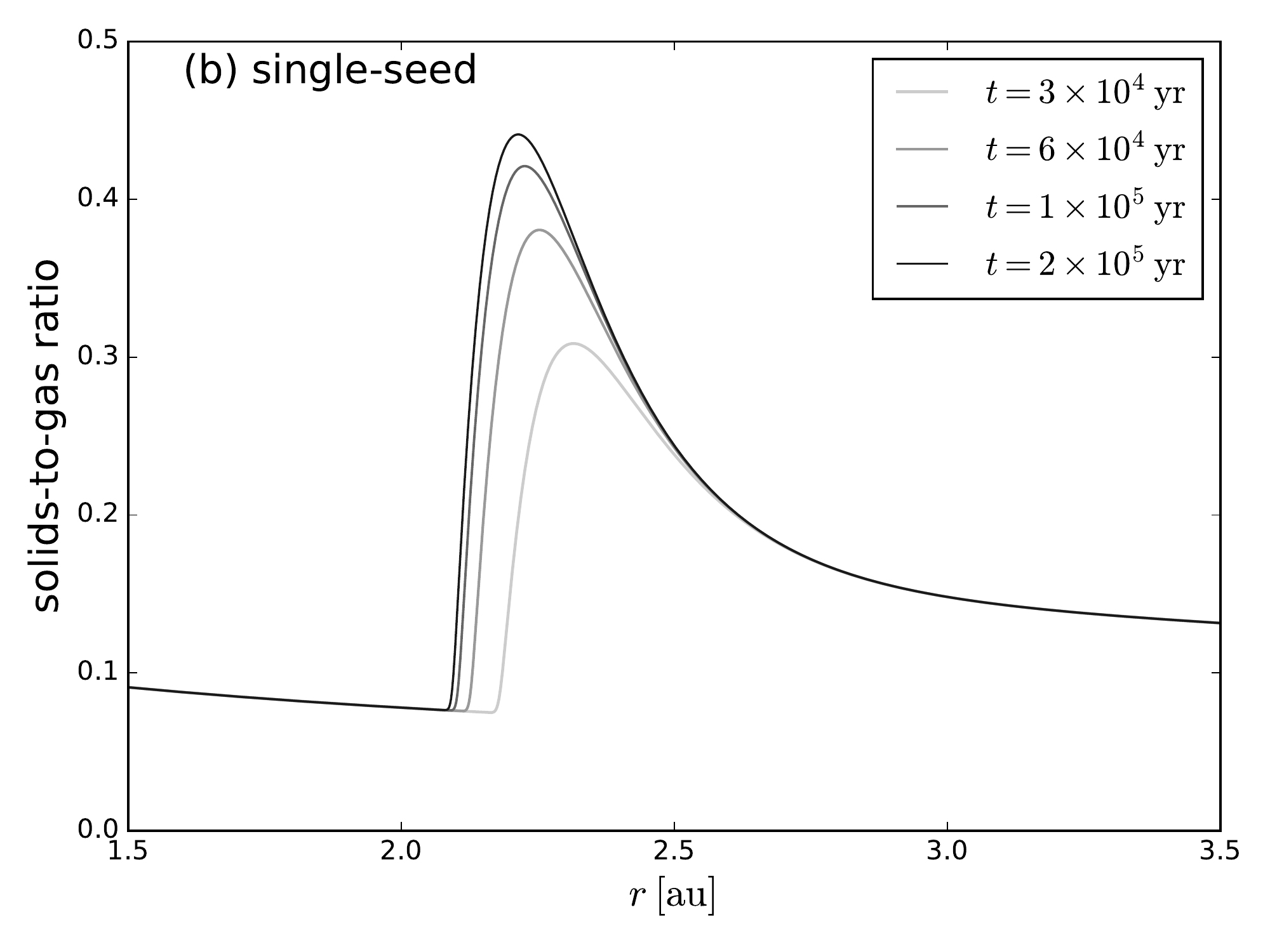}
		\includegraphics[width=0.495\textwidth]{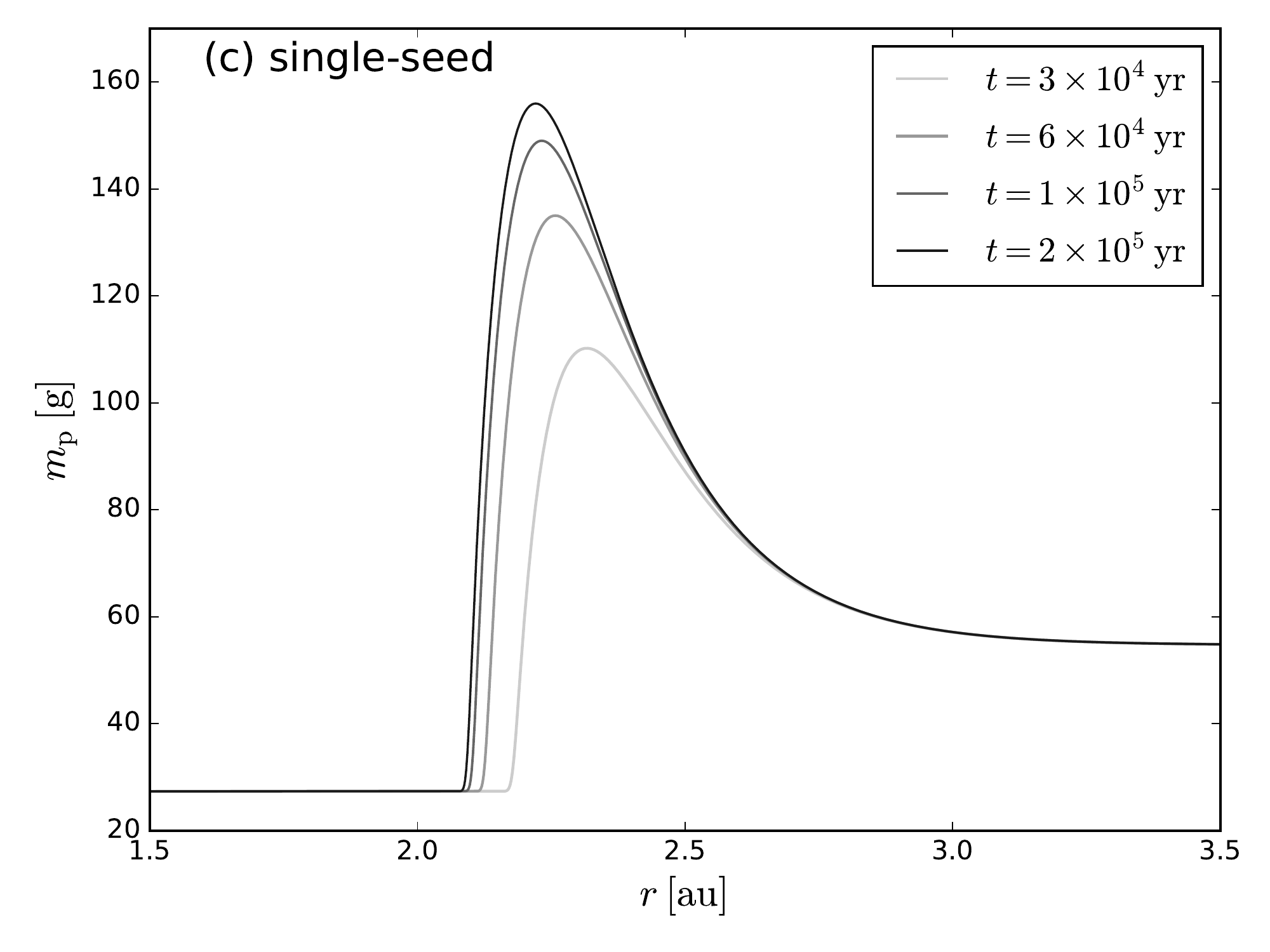}
		\includegraphics[width=0.495\textwidth]{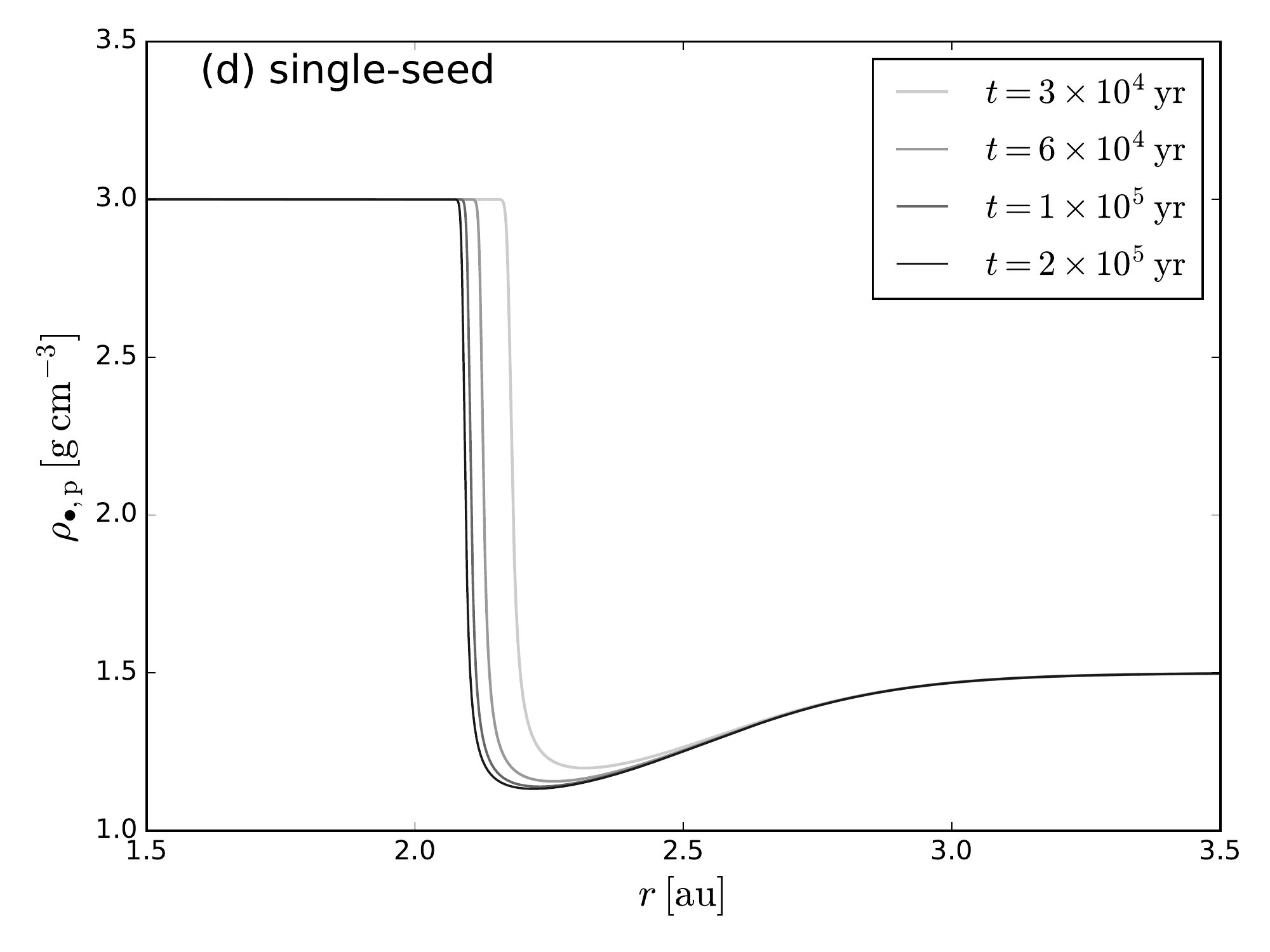}	
        \caption{The solution to the transport equations at different points in time, for the fiducial model parameters listed in Table \ref{table:fiducial}. At $t = 10^{5} \: \rm{yr}$, approximately 90\% of the steady-state peak in the ice surface density has formed. 
{\bf (a)} Surface densities $\Sigma$ of ice (solid lines) and vapor (dashed lines). The dotted line corresponds to the steady-state advection-only vapor surface density profile. {\bf (b)} Midplane pebbles-to-gas ratio $\rho_{\rm{peb}} / \rho_{\rm{gas}}$. {\bf (c)} Typical pebble mass $m_{\rm{p}}$. {\bf (d)} Typical pebble internal density $\rho_{\bullet,\rm{p}}$.\label{fig:timedependent-3snapshots}}
\end{figure*}

\section{Numerical implementation}
We make use of the open source partial differential equations solver \texttt{FiPy} \citep{FiPy:2009} to solve the transport equations numerically. The coupled systems of equations are solved on a cylindrical 1D grid that ranges from 1 to 5 au. 

We implement boundary conditions as follows. At the outer boundary $r_2 = 5$~au, the pebble mass flux $\mathcal{F}_{\rm{s/g}}\dot{M}_{\rm{gas}}$ is fixed\footnote{We do not fix the pebble surface density: for high values of $\alpha$, outward diffusion can be strong enough to enhance the surface density of pebbles at the outer boundary ($\Sigma_{\rm{peb, r2}}$) with respect to the expected value $\Sigma_{\rm{peb, r2}} = \dot{M}_{\rm{peb}} / (2 \pi r_{2} v_{\rm{peb}} (r_{2}))$ in case of no diffusion.}. We convert the input parameter $\tau_{3}$ to a physical pebble size $s_{\rm{p, start}}$ using \eq{stoppingtime}, and since our model does not include coagulation, we set the pebble size (which can be converted to a pebble mass $m_{\rm{p, start}}$) at the outer boundary $r_2$ to this value. For the single-seed model, the mass of the bare silicate core is then given by $m_{\rm{core}} = \zeta m_{\rm{p, start}} = m_{\rm{p, start}} / 2$. At the inner boundary $r_1 = 1$~au, we implement outflow boundary conditions: solid particles and water vapor are accreted to the central star at a constant rate. We also demand that $\Sigma_{\rm{ice}} = 0$ at $r_1$ and $\Sigma_{Z} = 0$ at $r_2$, for obvious reasons. Additionally, in the many-seeds implementation we have $\Sigma_{\rm{dirt}} = \Sigma_{\rm{ice}} = 0$ at $r_1$ and $\Sigma_{\rm{sil}} = \Sigma_{Z}= 0$ at $r_2$.

\begin{figure}[t]
	\centering
		\includegraphics[width=0.5\textwidth]{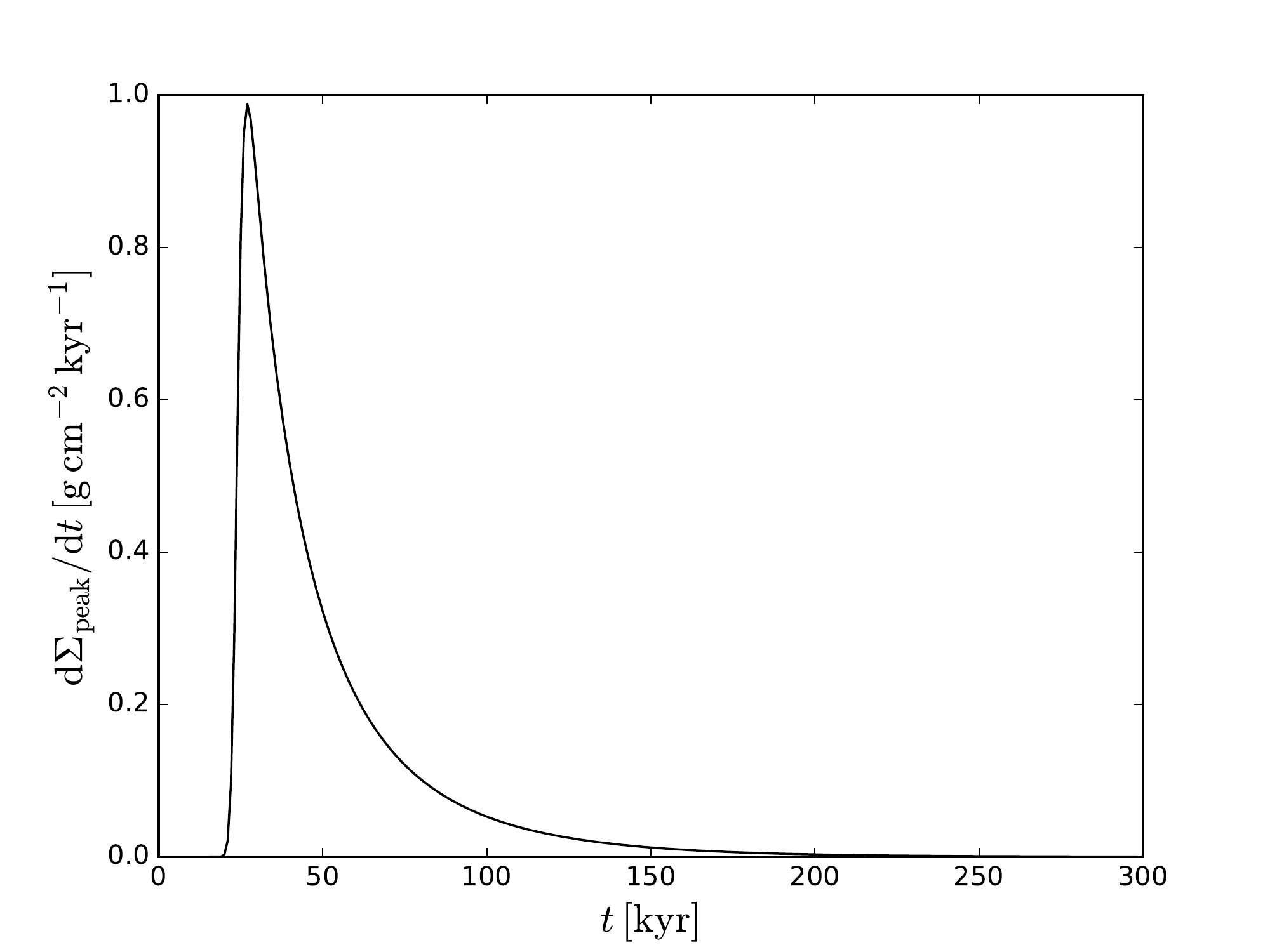}
\caption{Rate of change of the ice surface density at the peak location, as function of time. The rate of change increases until the icy pebble front has reached the peak location, after which the peak continues to grow at an exponentially decreasing rate, while the system is converging to its steady-state solution.\label{fig:timedependent-dpeakdt}}
\end{figure}

\subsection{Time-dependent}
In our time-dependent method, we adopt small time steps ($\Delta t \sim 100 \: \rm{yr}$) and a small value for the allowed residuals, to ensure mass conservation by demanding the residuals\footnote{\texttt{FiPy} linearizes the equations while our system is nonlinear; the residuals quantify this mismatch.} to be small. This method requires a lot of computational power because the system of partial differential equations is numerically very stiff,
but the advantage is that we find not only the steady-state solution to the equations, but are able to follow the solution in time. This allows us to get a measure of the time needed for the solution to settle into a steady-state. We present the time-dependent results for the simple, single-seed, model with fiducial input parameters in \se{timedependent}.

\subsection{Time-independent\label{sec:method-tin}}
Our time-independent method solves for the steady-state solution directly, by taking very large time-steps (in total 40 time steps that increase to $10^{7} \: \rm{yr}$) and decreasing the residuals threshold after each run, taking the solution from the previous run as the new input. We continue this process until we reach convergence. We present the steady-state results for the `simple, single-seed', `simple, many-seeds', `complete, single-seed', and `complete, many-seeds' models with fiducial input parameters in \se{results-fid-steadystate}.


\begin{table}
\caption{Fiducial model parameters}
\label{table:fiducial}
\centering
\begin{tabular}{l l l}
\hline
Gas accretion rate & $\dot{M}_{\rm{gas}}$ &$10^{-8} \: \rm{M}_{\odot} \rm{yr}^{-1}$\\
Pebbles-to-gas accretion rate & $\mathcal{F}_\mathrm{s/g}$ &0.8\\
Turbulence strength & $\alpha$ &$3 \times 10^{-3}$\\
Dimensionless stopping time at 3 au & $\tau_3$ &$3 \times 10^{-2}$\\
\hline
\end{tabular}
\tablefoot{See Sect.~2.7. $\tau_3$ translates into a physical pebble size at 3 au for the zero-model.}
\end{table}

\begin{figure*}[t]
	\centering
		\includegraphics[width=0.495\textwidth]{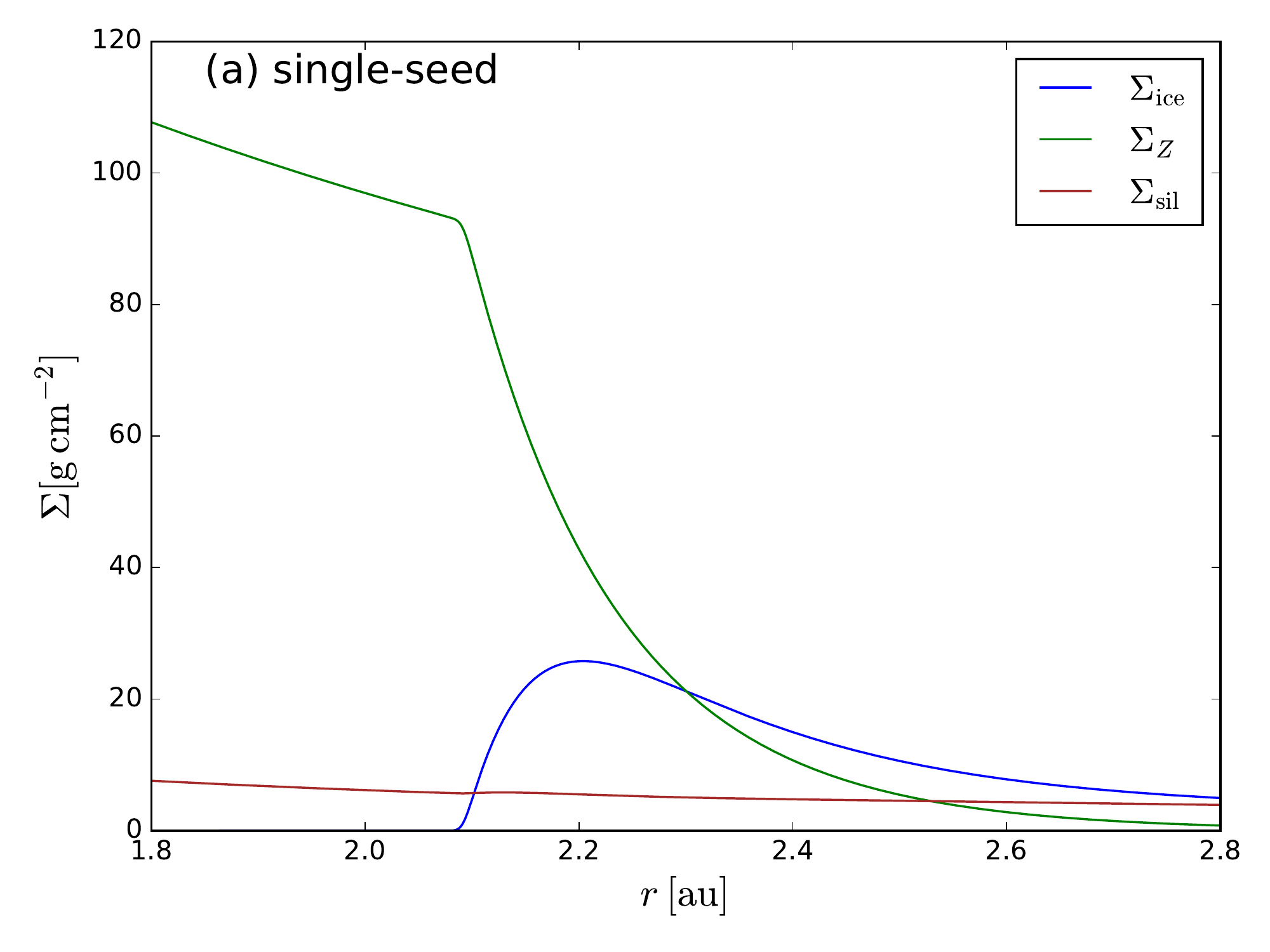}
		\includegraphics[width=0.495\textwidth]{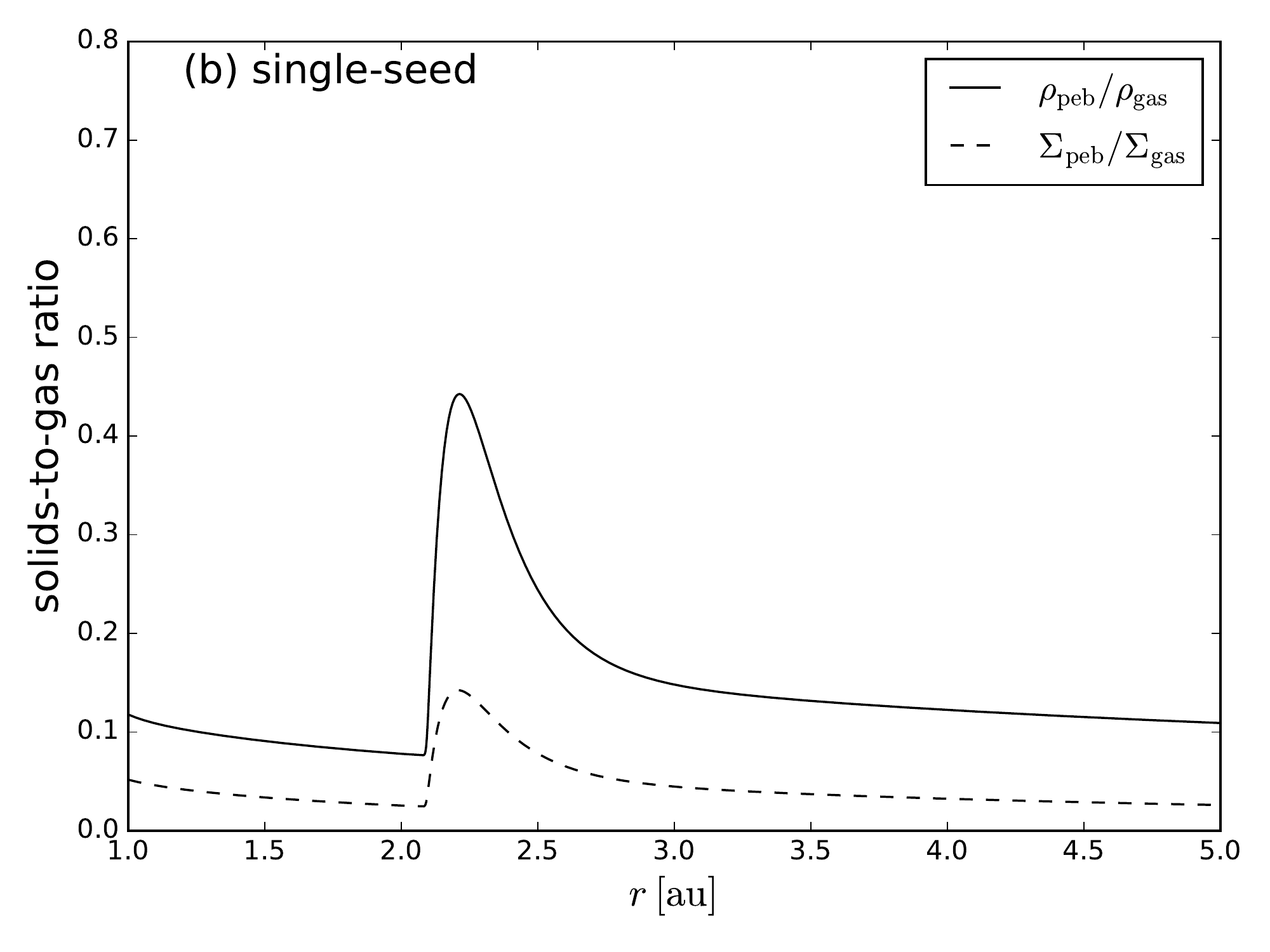}
		\includegraphics[width=0.495\textwidth]{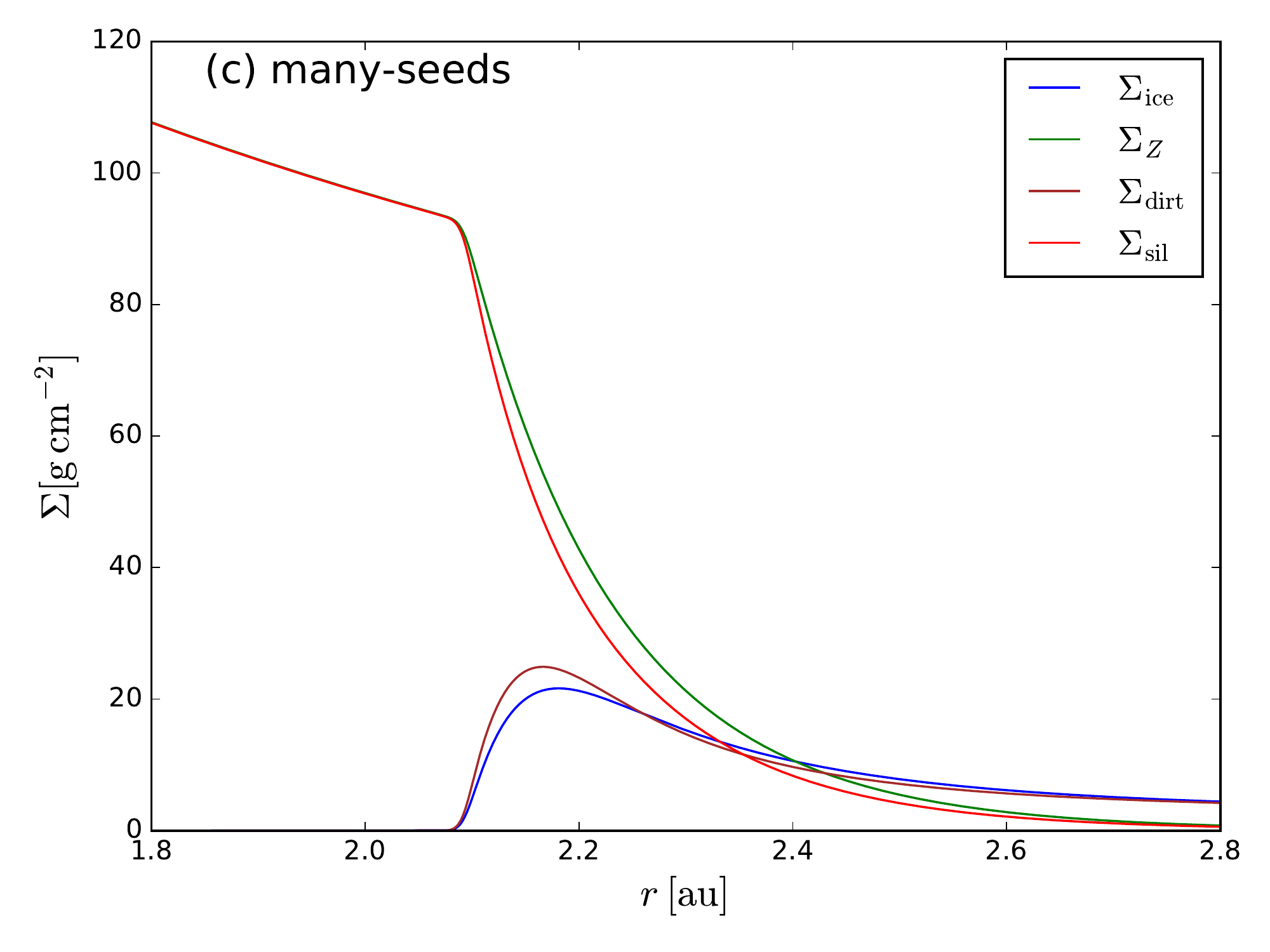}
		\includegraphics[width=0.495\textwidth]{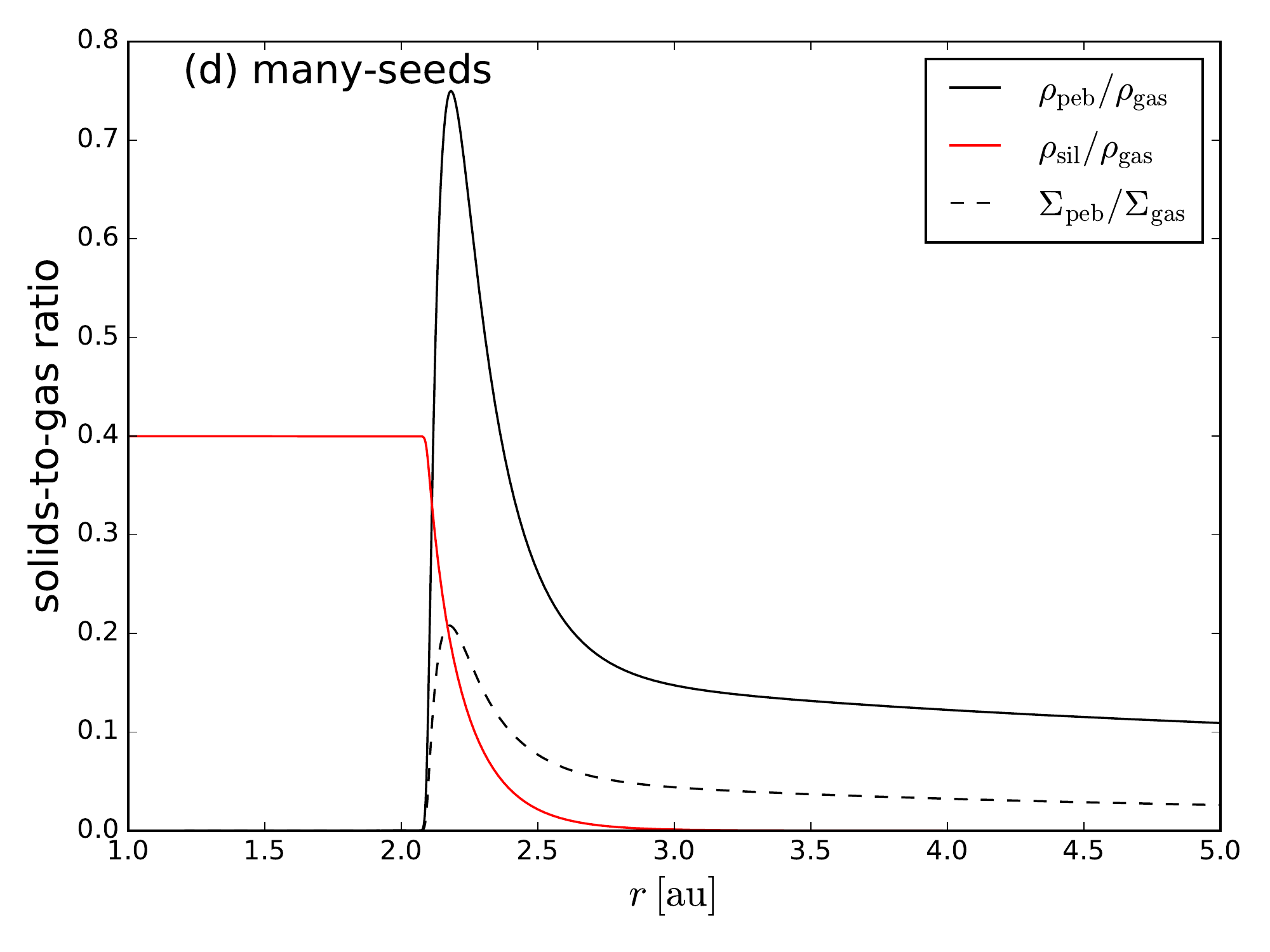}
        \caption{Steady-state results for the fiducial parameters, for the simple model with single-seed evaporation (upper two panels) and many-seeds evaporation (lower two panels). {\bf (a)} Surface densities of ice, silicates and vapor ($\Sigma_{\rm{ice}}$, $\Sigma_{\rm{sil}}$, and $\Sigma_{Z}$, respectively) for the single-seed model. The silicates are locked in the icy pebbles outside the snowline, whereas they are bare grains interior to the snowline. {\bf (b)} Midplane pebbles-to-gas ratio $\rho_{\rm{peb}} / \rho_{\rm{gas}}$, and pebbles-to-gas column density ratio $\Sigma_{\rm{peb}} / \Sigma_{\rm{gas}}$ for the single-seed model. {\bf (c)} Same as panel (a) but for the many-seeds model. The silicates are now divided in two populations: silicates that are locked up in icy pebbles ($\Sigma_{\rm{dirt}}$), and free silicates ($\Sigma_{\rm{sil}}$). Note that in panels (a) and (c) the $x$-axis ranges between $r = 1.8 - 2.8 \: \rm{au}$, for clarity. {\bf (d)} Same as (b), but for the many-seeds model and with an additional line for the midplane `free silicates'-to-gas ratio $\rho_{\rm{sil}} / \rho_{\rm{gas}}$.\label{fig:ss-fid-clean}}
\end{figure*}

\section{Results --- Fiducial model}\label{sec:results-fid}
In this section we introduce a standard set of parameters which we then use to illustrate different aspects of the model. The fiducial model is defined by the parameters listed in Table \ref{table:fiducial}. We first discuss the time-dependent results for the simple, single-seed, fiducial model in \se{timedependent}. We then go on to discuss the steady-state results for the other variants of the fiducial model in \se{results-fid-steadystate}. 

\subsection{Time-dependent solution}\label{sec:timedependent}
In \fg{timedependent-3snapshots} we plot several quantities as function of radial distance from the star $r$, at four points in time. In \fg{timedependent-3snapshots}a, the surface densities of ice (solid lines) and vapor (dashed lines) are plotted. Initially ($t=0$), the ice and vapor surface density are zero, $\Sigma_\mathrm{ice}=\Sigma_Z=0$. Icy pebbles are drifting in from the outer disk (right) to the inner disk (left). At the point where $\dot{\Sigma}_{Z,\mathrm{C/E}}>0$ the icy component of the pebbles evaporate. At the \textit{snowline} $r_\mathrm{snow}\approx2.1$ au all the H$_2$O is in the gas phase. Since the vapor is advected with the gas velocity $v_\mathrm{gas}$, which is much smaller than the pebble velocity $v_\mathrm{peb}$, the vapor surface density (dashed line) quickly exceeds that of the ice (solid line). The vapor density increases until the steady-state vapor distribution $\Sigma_{Z,\rm{a}}$ has been reached. $\Sigma_{Z,\rm{a}}$ is given by\footnote{Diffusion does not play a role in the steady-state, advection-only vapor surface density profile, because vapor has the same velocity as the gas and therefore the vapor concentration is independent of $r$.}:
\begin{equation}
    \label{eq:sigZan}
    \Sigma_{Z,\rm{a}} = \frac{\dot{M}_{\rm{gas}}}{3 \pi \nu}
\end{equation}
and is plotted by the dotted curve. Meanwhile, due to the outward diffusion of water vapor across the snowline and subsequent condensation onto the inward-drifting pebbles, the ice surface density just exterior to the snowline increases. Diffusion of vapor therefore results in a distinct bump in the ice profile and we denote the location corresponding to maximum $\Sigma_\mathrm{ice}$ the \textit{peak radius} $r_\mathrm{peak}$. A similar maximum is seen in the solids-to-gas ratio (\fg{timedependent-3snapshots}b), which rises to $\sim$$0.45$. After $t=2\times10^5$ yr a steady state has been reached.

The enhancement of the pebbles-to-gas ratio is reflected by the typical pebble mass $m_{\rm{p}}$ and density $\rho_{\bullet,\rm{p}}$, plotted in  \fg{timedependent-3snapshots}c and \fg{timedependent-3snapshots}d, respectively.
Pebbles at the inner boundary are half as massive as pebbles at the outer boundary, because at the inner boundary all the ice has evaporated off the pebbles, and the dust fraction of pebbles $\zeta = 0.5$. The typical pebble mass just outside the snowline increases over time, indicating that water vapor condenses onto the pebbles. This is also shown by the behaviour of the typical pebble density $\rho_{\bullet,\rm{p}}$. At the outer boundary, $\rho_{\bullet,\rm{p}} = 1.5 \: \rm{g} \: \rm{cm}^{-3}$, corresponding to pebbles containing as much water ice as silicates in mass; at the inner boundary, $\rho_{\bullet,\rm{p}} = 3 \: \rm{g} \: \rm{cm}^{-3}$, corresponding to pure silicate pebbles; whereas just outside the snowline, $\rho_{\bullet,\rm{p}}$ decreases over time to values close to unity -- corresponding to pure water ice pebbles. A decrease in particle internal density outside the snowline due to water condensation was also found by \citet{EstradaEtal2016}.

In \fg{timedependent-dpeakdt} we plot the growth rate of the peak in $\Sigma_{\rm{ice}}$ outside the snowline as function of time $t$. At the start of the simulation, ice is not yet present at the location of the peak. After the icy pebbles have reached the location of the eventual peak, at $t \sim 2 \times 10^{4} \: \rm{yr}$, the rate of growth decreases exponentially -- the peak continues to grow, but at an ever slower rate, while the system is converging to its steady-state solution.
The $e$-folding growth timescale is $\sim$$3 \times 10^{4} \: \rm{yr}$, which is approximately equal to the gas advection timescale $(r_{\rm{peak}} - r_{\rm{snow}}) / v_{\rm{gas}}$;  the time it takes for the vapor to traverse the peak and contribute to the steep vapor concentration gradient along which outward diffusion can take place. We will come back to the peak formation timescale in \se{results-parstudy}. The time in which 90\% of the peak forms is $\sim$$10^{5} \: \rm{yr}$.
Note that we have assumed that the incoming pebble mass flux does not change over time ($\mathcal{F}_{s/g}$ is constant at the outer boundary $r_2$). If the pebble flux would decrease on a timescale shorter than the time it takes to form the peak, a steady-state would not exist. If the pebble flux would decrease on a timescale longer than it takes to form the peak, we would reach a quasi-steady-state solution; in that case our model is fully applicable.

\subsection{Steady-state solution}\label{sec:results-fid-steadystate}
The time-dependent solution described in \se{timedependent} converges to the steady-state solution. In this section we take a closer look at the steady-state solution for the fiducial parameters, for the model designs discussed in \se{model}.

\subsubsection{Single-seed versus many-seeds}
In \fg{ss-fid-clean} we compare the simple single-seed model (upper two panels) with the simple many-seeds model (lower two panels). The left panels show the surface densities of ice, vapor, and in the many-seeds case, also of locked silicates (dirt) and free silicates. The ice profiles are very similar in both models, but in the many-seeds case there is an additional peak in the dirt surface density. This is because the free silicates behave like vapor: they are released throughout the evaporation front, and follow the gas profile interior to the snowline. Turbulent diffusion not only causes vapor to be transported outward and condense onto the pebbles, but leads to the same effect for the small silicates -- they diffuse outward and stick onto the pebbles, resulting in a peak in the dirt surface density distribution.

In the right panels we plot pebbles-to-gas midplane density ratios $\rho_{\rm{peb}} / \rho_{\rm{gas}}$ and column density ratios $\Sigma_{\rm{peb}} / \Sigma_{\rm{gas}}$. In the many-seeds model, the pebbles-to-gas ratio is the sum of the dirt-to-gas ratio and the ice-to-gas ratio. In both models, a clear peak is present just outside the snowline. The midplane density ratio is larger than the column density ratio due to settling of pebbles. In the many-seeds case, we also plot the free silicates-to-gas midplane density ratio $\rho_{\rm{sil}} / \rho_{\rm{gas}}$. This ratio is constant interior to the snowline because the silicates are perfectly coupled to the gas. 
In the many-seeds model, small silicate grains diffuse outward across the snowline, whereas in the single-seed model, there is no significant outward diffusion of the larger silicate pebbles due to the absence of a steep concentration gradient.  
Therefore, the peak in the pebbles-to-gas ratio outside the snowline is larger in the many-seeds model than in the single-seed model.

\subsubsection{Simple versus complete models}
\Fg{ss-fid-cleanvscomplete} shows the results for the complete model with the single-seed pebble design (upper two panels, a and b) and with the many-seeds pebble design (lower two panels, c and d). In \fg{ss-fid-cleanvscomplete}a (single-seed) and \fg{ss-fid-cleanvscomplete}c (many-seeds), we plot the surface densities of ice, vapor, and in the many-seeds case the surface densities of locked silicates (dirt) and free silicates (sil). Dashed lines denote the simple model results of \fg{ss-fid-clean}a and \fg{ss-fid-clean}c, for comparison. The grey lines correspond to the mean molecular weight of the gas. From the outer disk to the inner disk, the mean molecular weight increases from 2.34 $m_{\rm{H}}$ to 3.11 $m_{\rm{H}}$ across the snowline. Clearly, the peak in the ice surface density is higher and broader for the complete model than for the simple model. This is because collective effects (which reduce the pebble drift velocity) outweigh the effect of the enhanced gas pressure gradient in the evaporation front (which enhances the pebble drift velocity).
 In the complete model, therefore, pebbles effectively have a smaller radial velocity than in the simple model, and therefore the resulting ice peak is higher and broader. This is also the case in the many-seeds implementation. The width of the peak strengthens the justification of the vertical mixing assumption (see below).  

In \fg{ss-fid-cleanvscomplete}b and \fg{ss-fid-cleanvscomplete}d we compare the midplane pebbles-to-gas ratios between the simple and the complete model. Collective effects help to boost the pebbles-to-gas ratio, for the reason outlined above.

\begin{figure*}[t]
	\centering
		\includegraphics[width=0.495\textwidth]{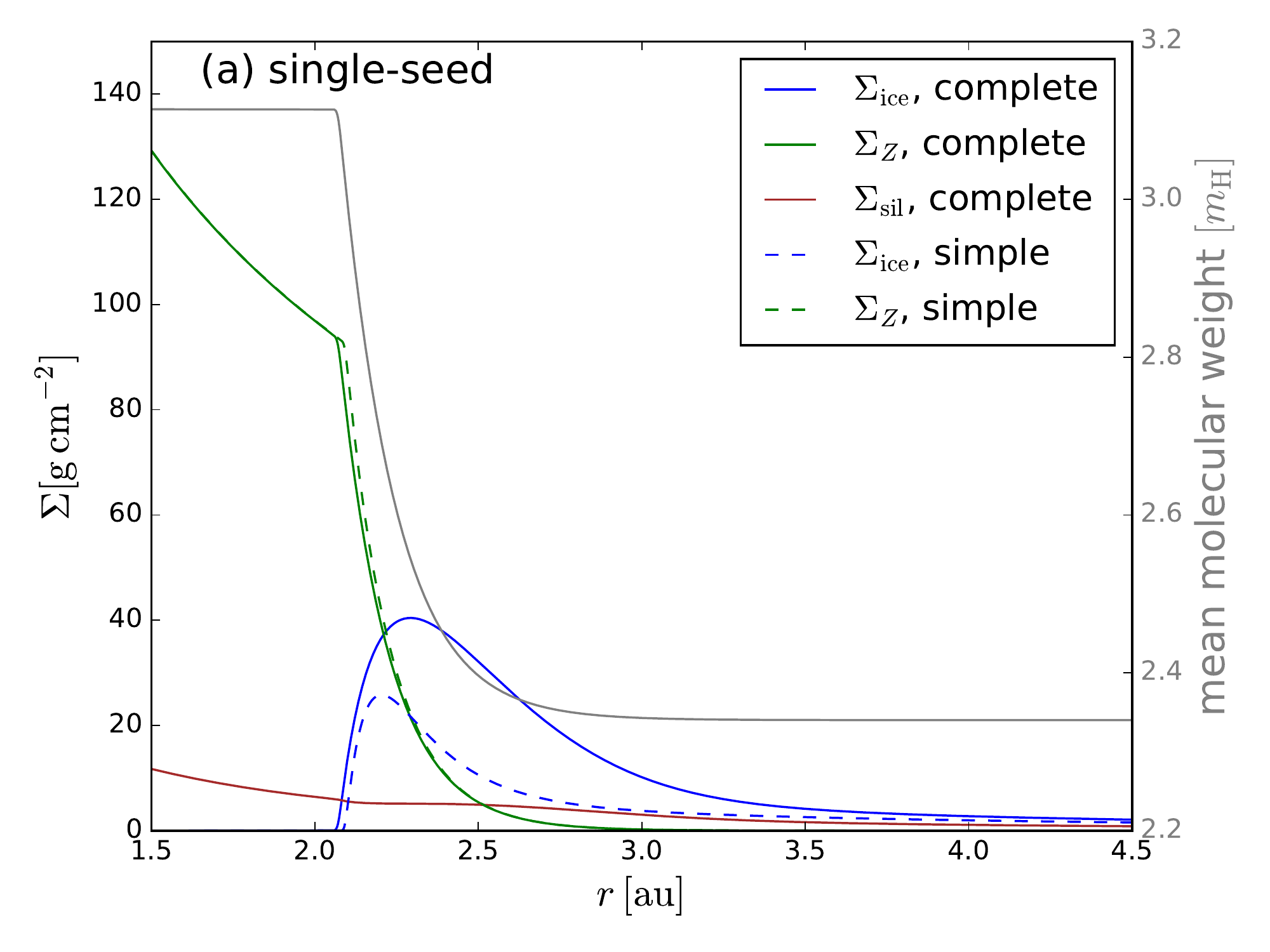}
		\includegraphics[width=0.495\textwidth]{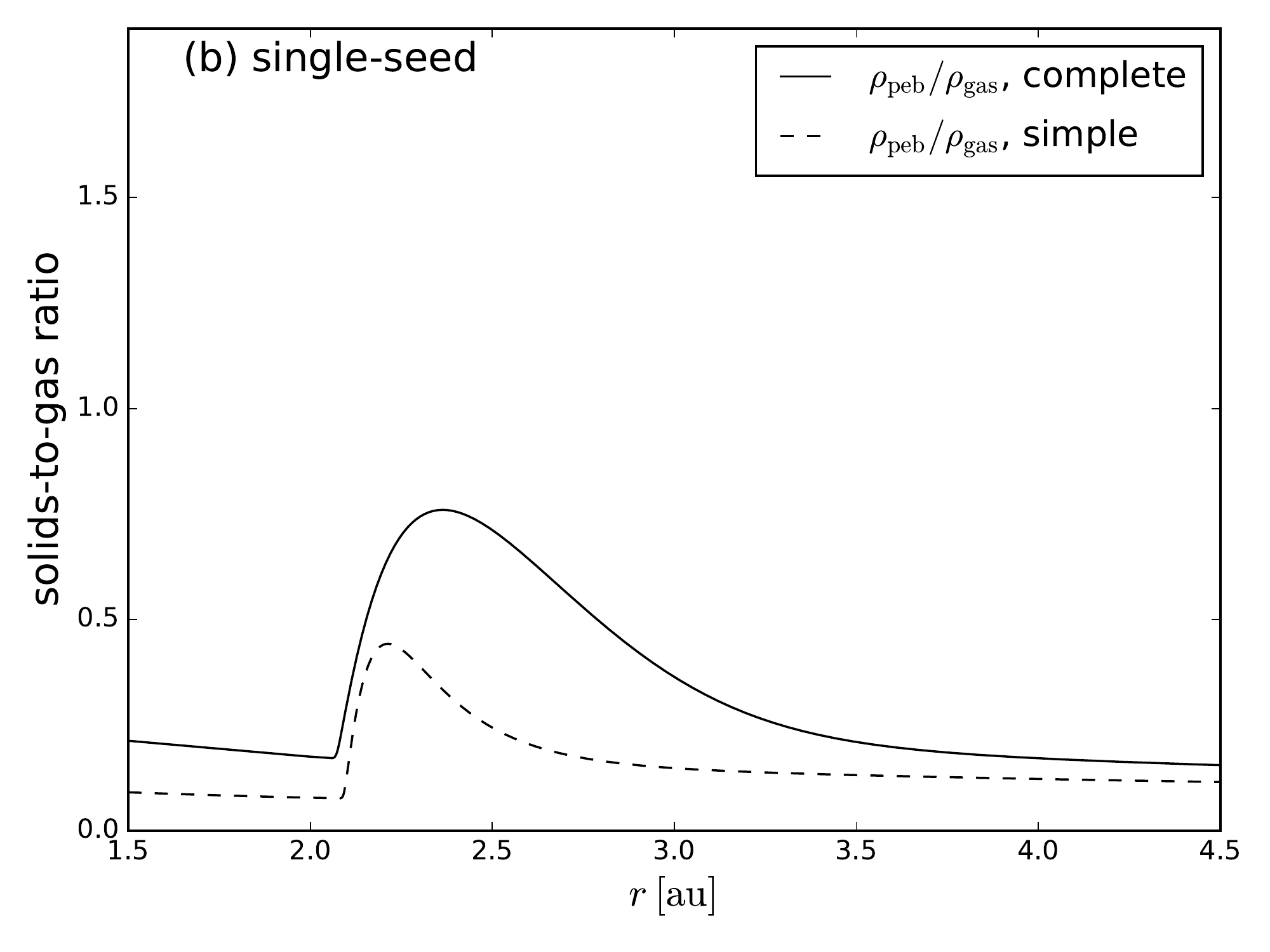}
		\includegraphics[width=0.495\textwidth]{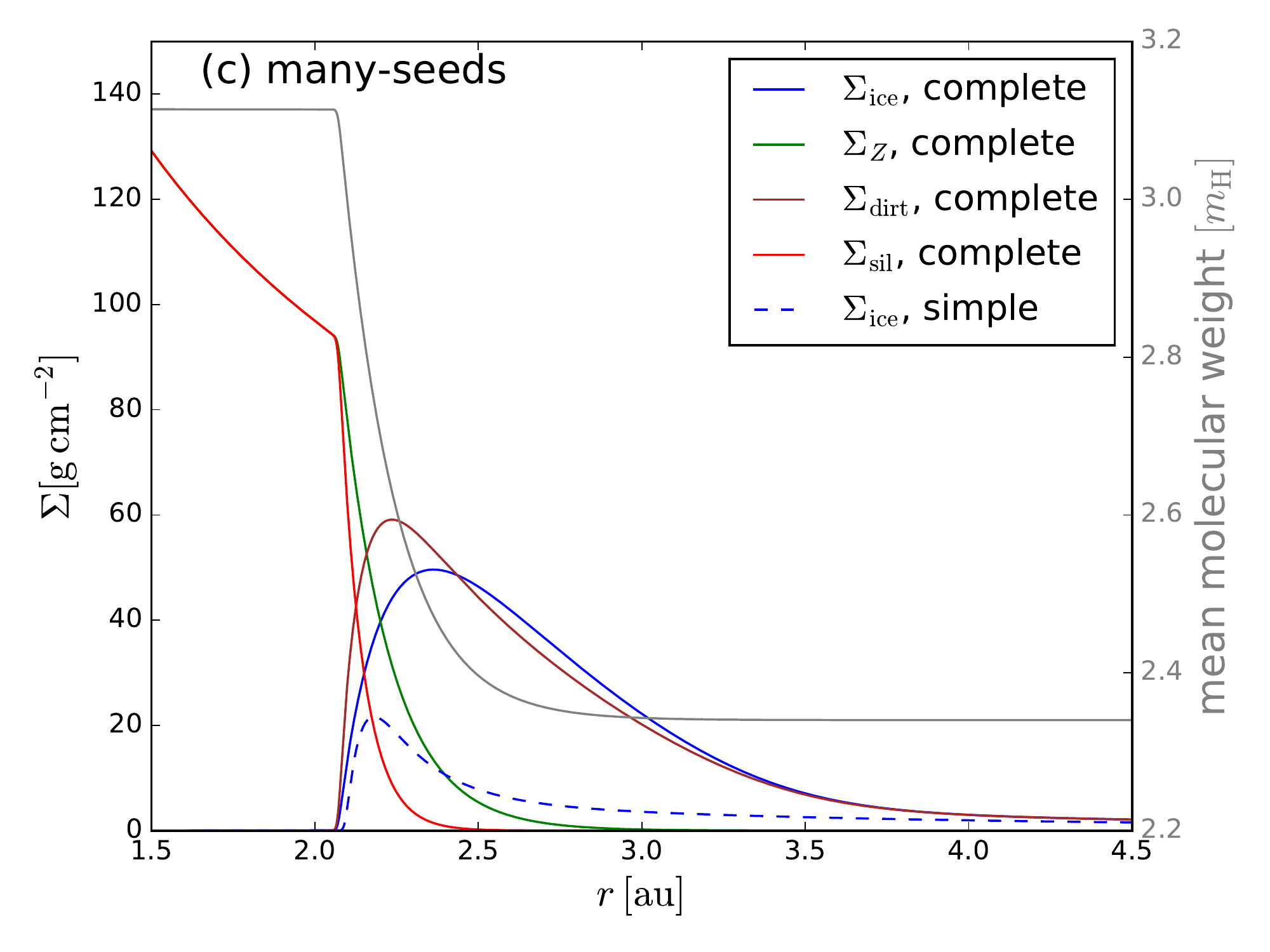}
		\includegraphics[width=0.495\textwidth]{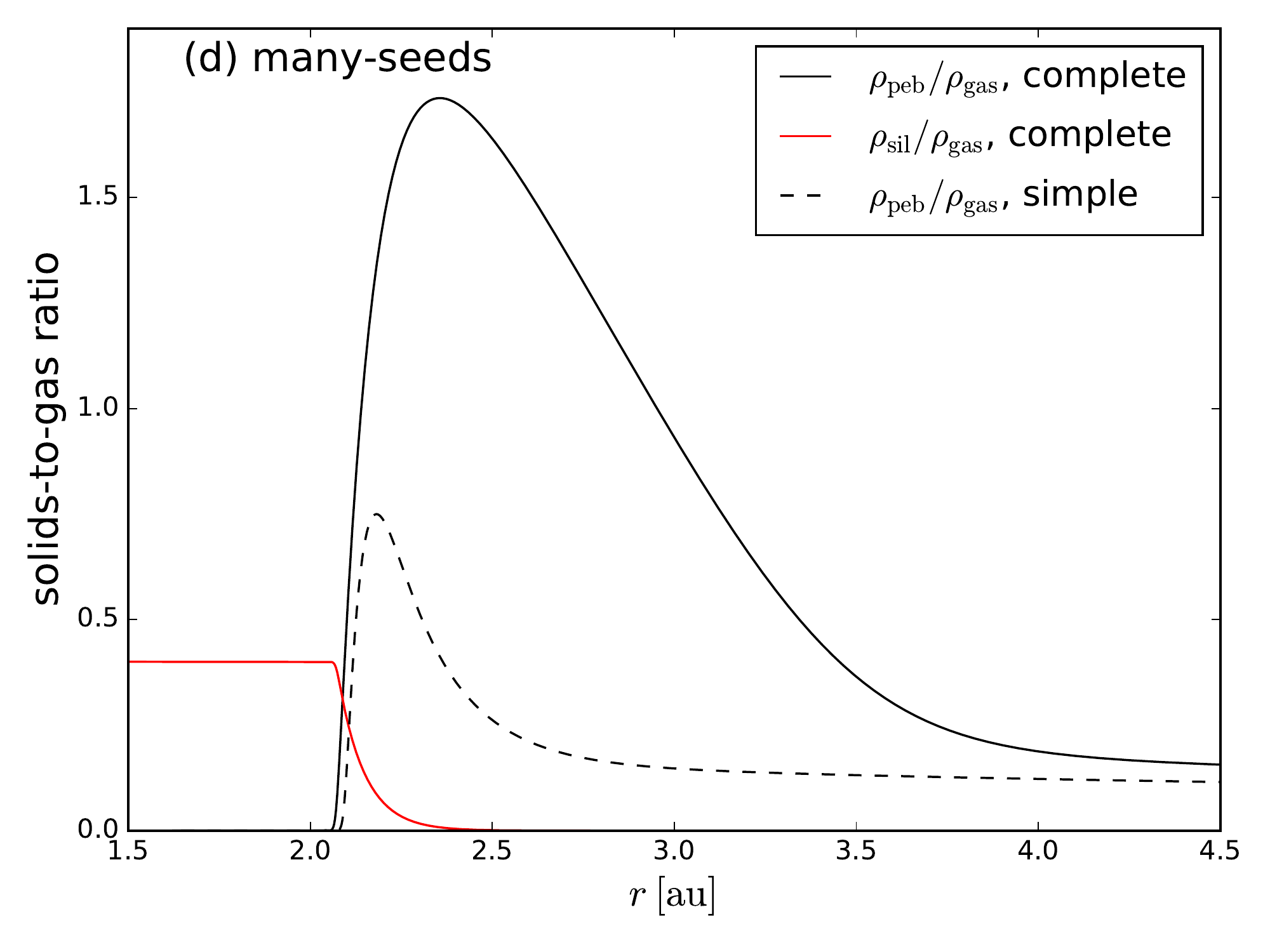}
        \caption{Comparison of steady-state results for the fiducial parameters between the simple model (dashed lines) and the complete model (solid lines), which includes collective effects and the effects of the variation of the mean molecular weight $\mu$. The upper two panels correspond to the single-seed evaporation model; the lower two panels correspond to the many-seeds evaporation model. In the latter case, we made the conservative choice of not including the free silicates in the back-reaction onto the gas. {\bf (a)} Surface densities of ice ($\Sigma_{\rm{ice}}$), vapor ($\Sigma_{Z}$) and silicates ($\Sigma_{\rm{sil}}$) for the single-seed model. The grey line indicates the mean molecular weight of the gas in units of proton mass $m_{\rm{H}}$. {\bf (b)} Midplane pebbles-to-gas ratio $\rho_{\rm{peb}} / \rho_{\rm{gas}}$. {\bf (c)} Same as (a), but the silicates are now divided into `free silicates' ($\Sigma_{\rm{sil}}$) and `locked silicates' ($\Sigma_{\rm{dirt}}$). For clarity, we only show the clean $\Sigma_{\rm{ice}}$ profile for comparison. {\bf (d)} Same as (b), with an additional line denoting the `free silicates'-to-gas ratio $\rho_{\rm{sil}} / \rho_{\rm{gas}}$.\label{fig:ss-fid-cleanvscomplete}}
\end{figure*}

\subsubsection{Validation of vertical mixing assumption}
As discussed in \se{assumptions}, we assume rapid vertical mixing of vapor. 
However, it could be argued that (part of) the released vapor recondenses onto icy pebbles before relaxing to the same vertical distribution as the gas, thereby boosting the condensation rate. This assumption also extends to the small silicate grains in the many-seeds model. Similarly, one could argue that the silicate grains can collide with a pebble before they can diffuse to higher vertical layers ({\it e.g.} \citet{2016ApJ...822..111K}). 
In order to investigate what the difference is in terms of the steady-state ice distribution between two extreme cases, we take the simple, single-seed, fiducial model and run it once with $H_{Z} = H_{\rm{gas}}$ and once with $H_{Z} = H_{\rm{peb}}$. We compare the steady-state outcomes of the two cases in \fg{Hcomparison}. Even though the location of the snowline and of the pebbles-to-gas peak differs a bit between the two cases, the height of the peak in the midplane pebbles-to-gas ratio increases only by $\sim$$0.3$\% if one takes $H_{Z} = H_{\rm{peb}}$ instead of $H_{Z} = H_{\rm{gas}}$. 
Similarly, assuming a smaller scale height for the silicate grains in the evaporation region would probably not significantly change the height of the peak in the `locked' silicates, since the silicate grains behave like vapor. In the rest of this paper, we assume $H_{Z} = H_{\rm{sil}} = H_{\rm{gas}}$.

\begin{figure*}[t]
	\centering
		\includegraphics[width=0.33\textwidth]{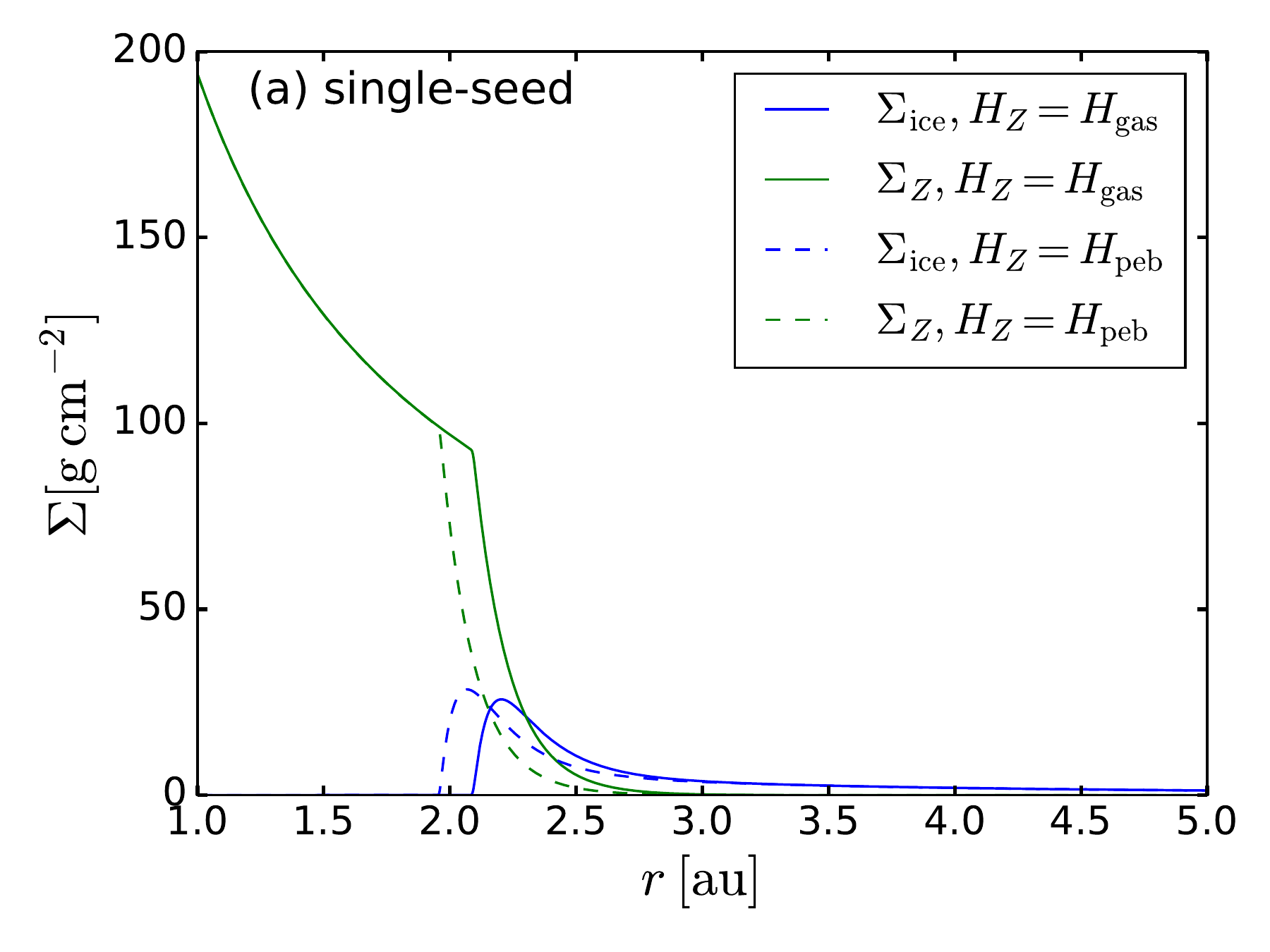}
		\includegraphics[width=0.33\textwidth]{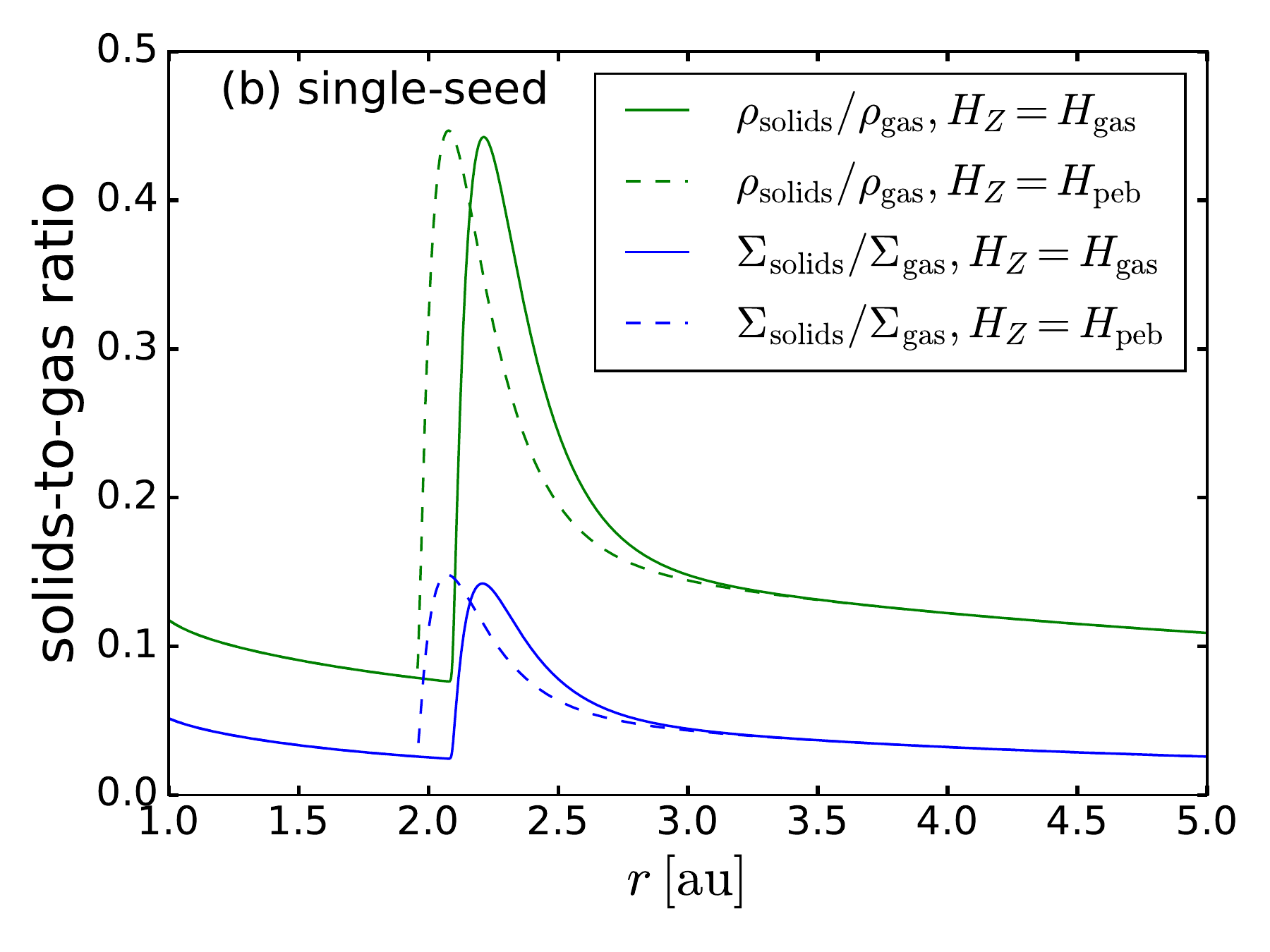}
		\includegraphics[width=0.33\textwidth]{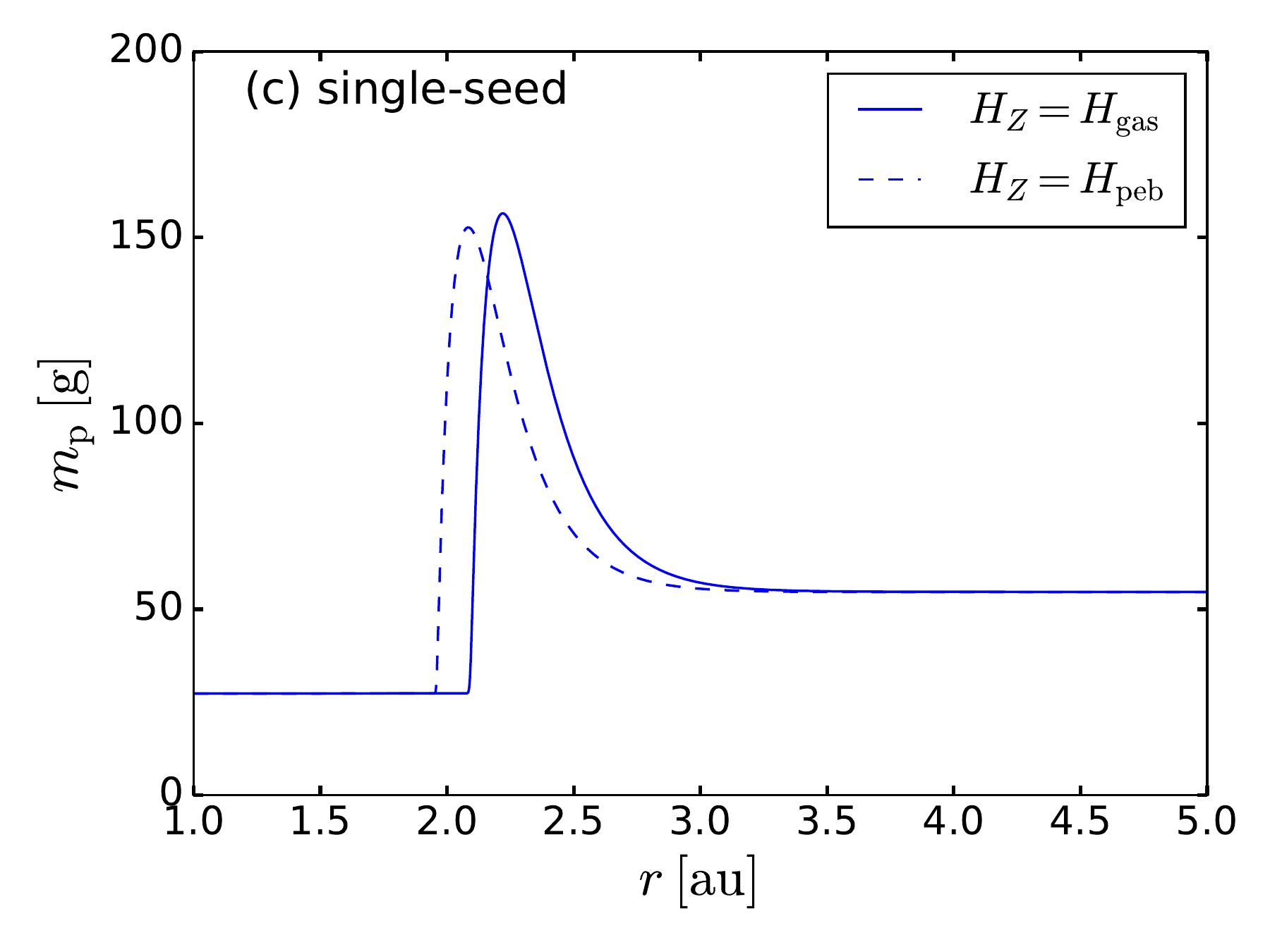}
        \caption{Comparison of steady-state results for the simple, single-seed, fiducial model between the case where vapor has the same vertical distribution as the background gas ($H_{Z} = H_{\rm{gas}}$) and the case where vapor has the same scale height as the pebbles ($H_{Z} = H_{\rm{peb}}$). In all three panels, solid lines correspond to the $H_{Z} = H_{\rm{gas}}$ case and dashed lines correspond to the $H_{Z} = H_{\rm{peb}}$ case. {\bf (a)} Steady-state water ice (blue) and water vapor (green) surface densities. {\bf (b)} Midplane solids-to-gas ratios (green) and solids-to-gas surface densities ratios (blue). {\bf (c)} Typical pebble mass $m_{\rm{p}}$.\label{fig:Hcomparison}}
\end{figure*}

\section{Streaming instability conditions}\label{sec:results-parstudy}

\begin{figure*}[t]
	\centering
		\includegraphics[width=0.495\textwidth]{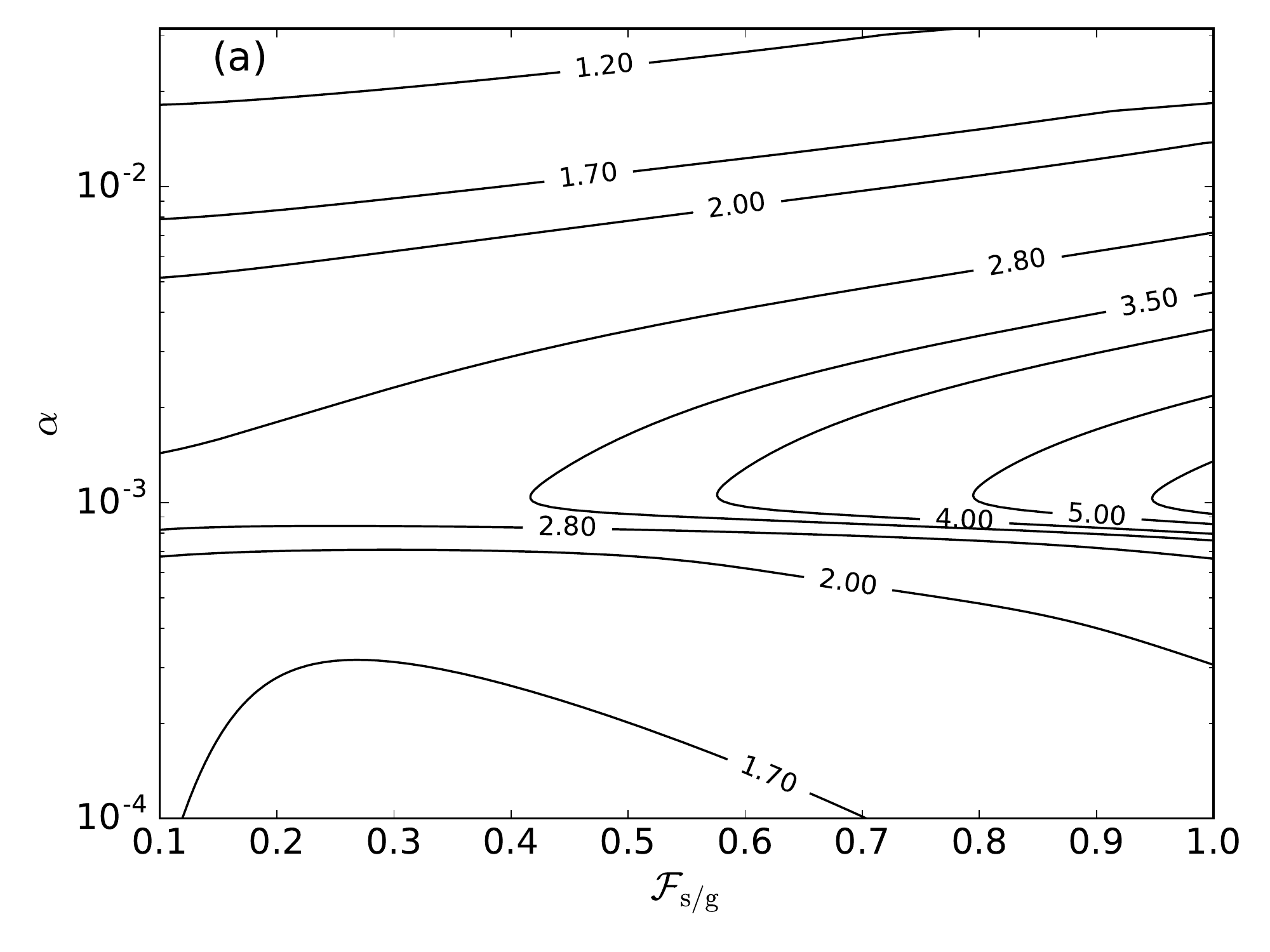}
		\includegraphics[width=0.495\textwidth]{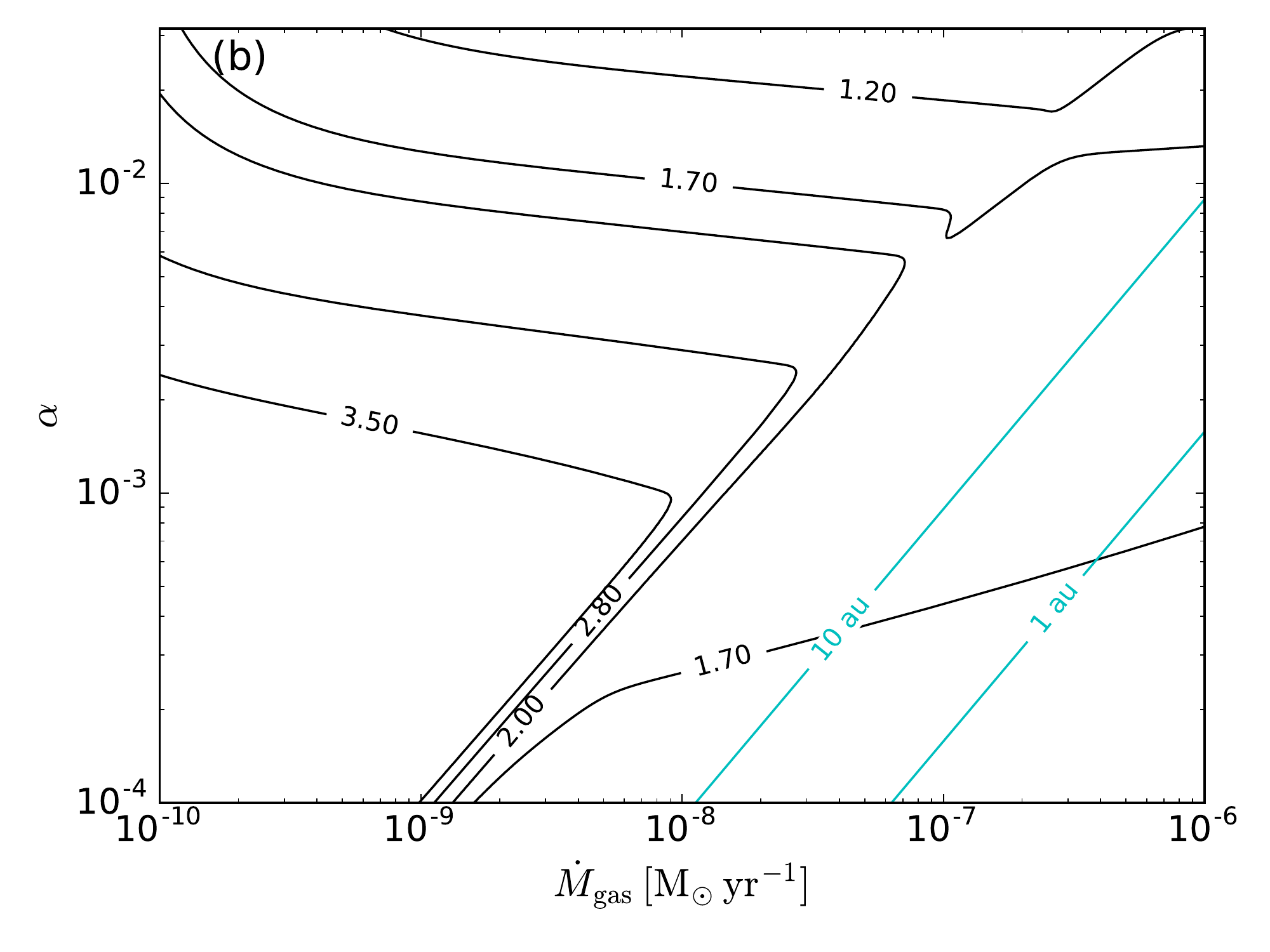}
        \caption{{\bf (a)} Contours of constant relative ice surface density peak heights $f_{\Sigma,\mathrm{peak}}$ (see text for details) as function of turbulence parameter $\alpha$ and solids-to-gas accretion rate $\mathcal{F}_{\rm{s/g}}$. The gas accretion rate $\dot{M}_{\rm{gas}}$ is fixed to $10^{-8} M_{\odot} \: \rm{yr}^{-1}$ and the initial size of the pebbles $\tau_{3}$ is fixed to $0.03$. {\bf (b)} Contours of constant relative ice surface density peak heights (see text for details) as function of the turbulence parameter $\alpha$ and the gas accretion rate $\dot{M}_{\rm{gas}}$. The solids-to-gas accretion rate ratio $\mathcal{F}_{\rm{s/g}}$ is fixed to 0.4 and the initial size of the pebbles $\tau_{3}$ is fixed to $0.03$. The cyan lines denote a Toomre parameter ($Q_{T}$) of unity at 10 au (1 au); the space below these lines has $Q_{T} < 1$ and is gravitationally unstable at 10 au (1 au).\label{fig:nml-contour-plots}}
\end{figure*}

\citet{2015A&A...579A..43C} have investigated for what values of the solids-to-gas column density ratio streaming instability occurs, as function of the stopping time of the solid particles. They found that, typically, the solids-to-gas column density ratio should exceed $\sim$$2\%$ for stopping times $\sim$$5 \times 10^{-2}$. However, these results should be interpreted with caution, since \citet{2015A&A...579A..43C} only accounted for self-driven (Kelvin-Helmholtz) turbulence. We do include global turbulence, and therefore cannot simply use their conditions for streaming instability. A more robust condition for streaming instability in the presence of global turbulence is probably a midplane solids-to-gas ratio that reaches order unity \citep{2007ApJ...662..627J}.

In this section we investigate under what disk conditions, the midplane solids-to-gas ratio near the snowline gets most enhanced due to the effect of water diffusion and condensation. To this end, we have constructed a semi-analytic approximate model to be able to quickly test for many different values of the input parameters (\se{keypars}), within the single-seed implementation (Sect. 2.4). Before presenting the results, we will provide a short summary of the semi-analytical model, which is discussed in greater detail in Appendix~C.

\subsection{Semi-analytical model: Summary}\label{sec:summarysam}
Our semi-analytical model is an analytic prescription to find the approximate location, height and width of the ice surface density at the location of the ice peak, $r_{\rm{peak}}$, of the steady-state (time-independent) solution (see Sect. 3.2 for our numerical method to find the steady-state solution to the transport equations). For simplicity, we have tailored the semi-analytical model towards the `single-seed' design. (An extension to also include the `many-seeds' design will be considered in a future work).

The model works by iteratively solving a set of analytical equations, until convergence is reached. The key parameter is the ratio between the pebble velocity at $r_{\rm{peak}}$ and the normalized diffusivity at this location, $\epsilon = v_{\rm{peb}} (r_{\rm{peak}}) / D(r_{\rm{peak}}) r_{\rm{peak}}$. Initially, we start with an estimate for $\epsilon$, which can, for example, be taken from the zero-model solution (Sect. 2.7). For a certain $\epsilon$ the semi-analytical model then updates the surface density at the peak ($\Sigma_{\rm{peak}}$) as well as $r_{\rm{peak}}$ (or rather the difference between $r_{\rm{ice}}$ and $r_{\rm{peak}}$). With these updates we obtain a new value of $\epsilon$ and the procedure can be repeated until the fractional change in the parameters has become sufficiently low.

We emphasize that the semi-analytical model gives approximate solutions (correct at the $\sim$$20\%$ level), but that they are very useful since they come at almost zero computational costs. It is therefore ideal to be applied to the parameter searches that we consider in this section. In Appendix~C the model is described in detail and compared to runs of the numerical steady-state model.

\subsection{Enhancement of the ice surface density}

We first look at the relative effect of enhancing the ice surface density near the snowline: for each disk model, we normalize the resulting peak ice surface density by the ice surface density at the peak radius in case there is only advection:
\begin{equation}
    f_{\Sigma,\mathrm{peak}} = \frac{\Sigma_{\rm{ice,peak}}}{(1-\zeta) \mathcal{F}_{\rm{s/g}} \dot{M}_{\rm{gas}} / 2 \pi r_{\rm{peak}} v_{\rm{peb}}}
\end{equation}
where $v_\mathrm{peb}$ is calculated in the absence of condensation and evaporation.
As before, we take $\zeta = 0.5$. In \fg{nml-contour-plots}a we plot contours of constant $f_{\Sigma,\rm{peak}}$ as function of $\alpha$ and $\mathcal{F}_{\rm{s/g}}$, where $\dot{M}_{\rm{gas}}$ is fixed at $10^{-8} M_{\odot} \: \rm{yr}^{-1}$ and $\tau_{3}$ is fixed at $0.03$. Clearly, the higher the solids-to-gas accretion rate $\mathcal{F}_{\rm{s/g}}$, the higher $ f_{\Sigma,\mathrm{peak}}$ -- at fixed $\alpha$ the normalized ice peak increases from low $\mathcal{F}_{\rm{s/g}}$-values to high $\mathcal{F}_{\rm{s/g}}$-values, due to collective effects playing an increasingly important role. The plot also shows that disks with $\alpha$-values of about $10^{-3}$ are best at enhancing the ice surface density with respect to the advection-only expected ice surface density. This is because higher values of $\alpha$ lead to a broader and relatively lower peak, while lower $\alpha$-values, indicating larger gas densities, move pebbles into the Stokes drag regime. The fact that the relative height of the ice peak becomes smaller when pebbles enter the Stokes regime, is explained by the pebble size-dependency of the drift velocity. In the Epstein regime the drift velocity depends linearly on the pebble size, whereas in the Stokes regime the drift velocity is proportional to the square of the pebble size. This means that when pebbles enter the Stokes regime, their stopping time -- and hence their drift velocity -- increases rapidly. The effect of this is that $f_{\Sigma,\rm{peak}}$ decreases rapidly. Therefore, the transition from the Epstein regime to the Stokes drag regime is characterized by closely-spaced contours.

\Fg{nml-contour-plots}b shows contours of constant $f_{\Sigma,\rm{peak}}$, as function of $\alpha$ and $\dot{M}_{\rm{gas}}$. $\mathcal{F}_{\rm{s/g}}$ is fixed at 0.4 and $\tau_3$ is fixed at 0.03. The cyan lines correspond to a Toomre parameter ($Q_T$) of unity at 10 au (1 au). $Q_T$ is defined as:
\begin{equation}
Q_T = \frac{c_s \Omega}{\pi G \Sigma_{\rm{gas}}}
\end{equation}
where $G$ is the gravitational constant. A disk is gravitationally unstable if $Q_T < 1$. In \fg{nml-contour-plots}b, models in the parameter space below the cyan line have $Q_T < 1$ at 10 au (1 au). In \fg{nml-contour-plots}b densely-spaced contours again indicate the transition between Epstein and Stokes drag.

The enhancement of the ice surface density near the snowline is relatively modest, but enrichment factors of a few can still be important for triggering streaming instability \citep{2009ApJ...704L..75J}. Besides, \fg{nml-contour-plots} corresponds to our single-seed model, which is conservative. Larger enhancements can be achieved in the framework of the many-seeds model (see \fg{ss-fid-clean}). We can imitate the many-seeds model in the analytical model of \se{toy-model}, crudely, when we let $\zeta\rightarrow0$, which assumes that micron-sized silicate grains behave like vapor. In that case we see $f_{\Sigma, \mathrm{peak}}$ increase by another factor of two (results not plotted). Therefore, total enhancements of $f_{\Sigma,\mathrm{peak}}\sim5$--$10$ are feasible, which significantly boost the likelihood of triggering streaming instability.

\begin{figure*}[t]
	\centering
		\includegraphics[width=0.495\textwidth]{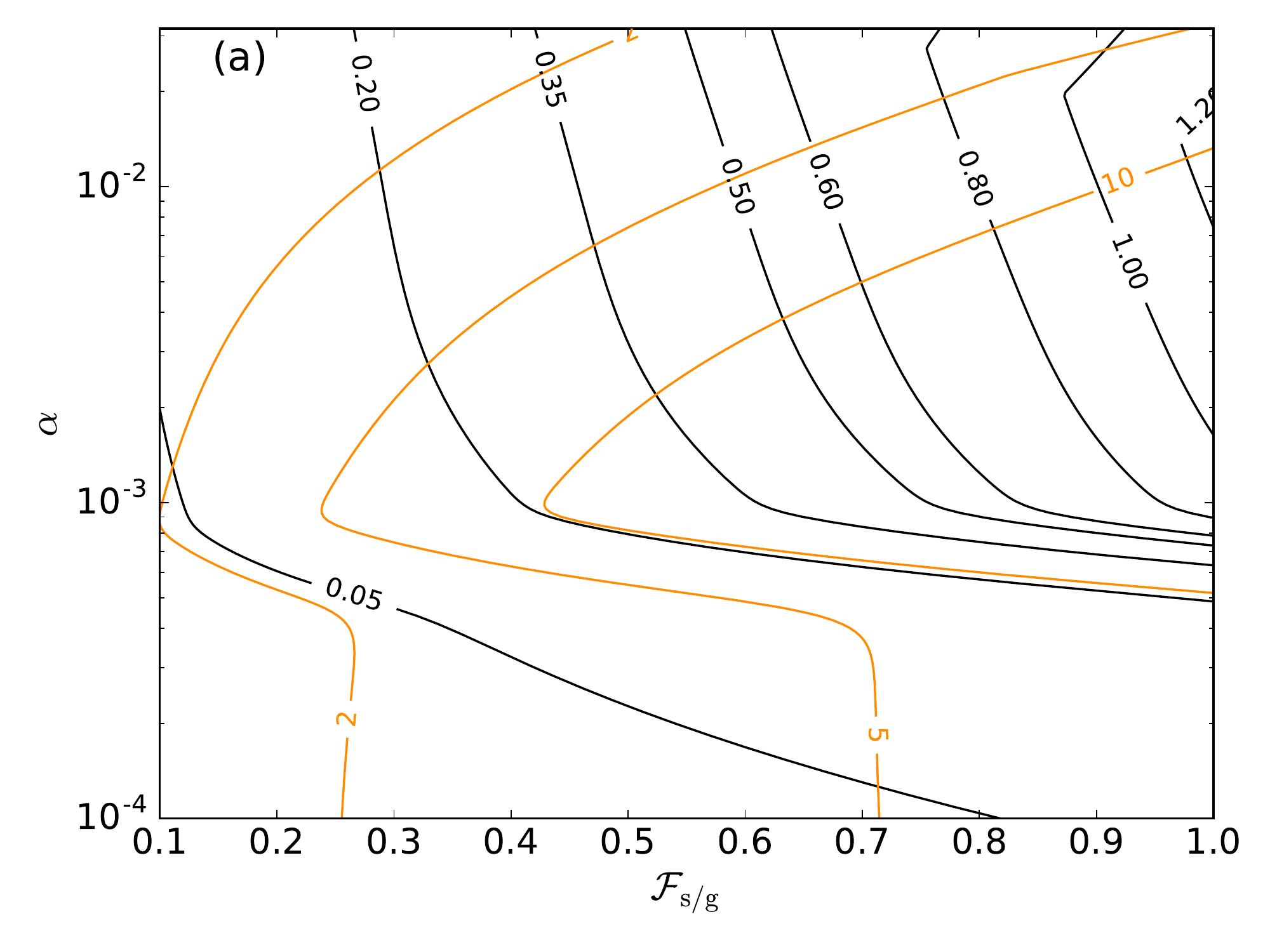}
		\includegraphics[width=0.495\textwidth]{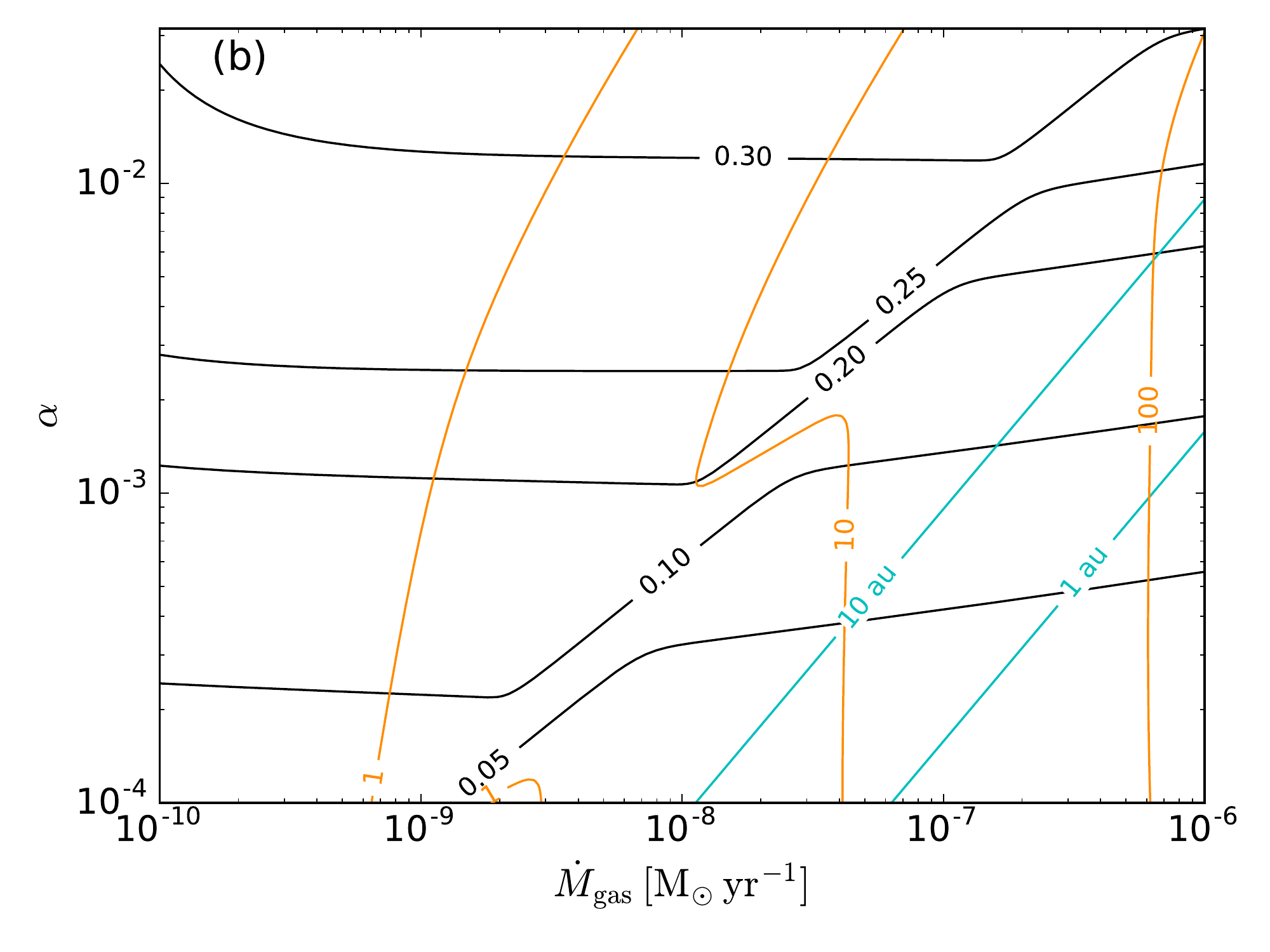}
        \caption{{\bf (a)} Contours of midplane solids-to-gas ratios (black) as function of the turbulence parameter $\alpha$ and the solids-to-gas accretion rate $\mathcal{F}_{\rm{s/g}}$. The gas accretion rate $\dot{M}_{\rm{gas}}$ is fixed to $10^{-8} M_{\odot} \: \rm{yr}^{-1}$ and the initial size of the pebbles $\tau_{3}$ is fixed to $0.03$. Orange contours denote the amount of solid material (in Earth masses) that is required to form the eventual peak in the ice surface density just outside the snowline. {\bf (b)} Contours of midplane solids-to-gas ratios as function of $\alpha$ and $\dot{M}_{\rm{gas}}$. The solids-to-gas accretion rate $\mathcal{F}_{\rm{s/g}}$ is fixed to 0.4 and the initial size of the pebbles $\tau_{3}$ is fixed to $0.03$. Orange contours denote the amount of solid material (in Earth masses) that is required to form the eventual peak in the ice surface density just outside the snowline. The cyan lines denote a Toomre parameter ($Q_{T}$) of unity at 10 au (1 au); the space below these lines has $Q_{T} < 1$ and is gravitationally unstable at 10 au (1 au).\label{fig:contour-plots}}
\end{figure*}

\subsection{Solids-to-gas ratios}\label{sec:siresults}

Let us now look at the absolute results, in terms of the midplane solids-to-gas ratio. In \fg{contour-plots}a, black lines denote contours of constant peak midplane solids-to-gas ratios, as function of $\alpha$ and $\mathcal{F}_{\rm{s/g}}$. We again fix $\dot{M}_{\rm{gas}} = 10^{-8} M_{\odot} \: \rm{yr}^{-1}$ and $\tau_{3} = 0.03$. The orange contours correspond to the total amount of solids in units of Earth masses that is needed to form the eventual peak. This quantity ($M_{\rm{solids,needed}}$) is calculated as:
\begin{equation}\label{eq:neededsolids}
M_{\rm{solids,needed}} \sim \frac{\mathcal{F}_{\rm{s/g}} \dot{M}_{\rm{gas}} (r_{\rm{peak}} - r_{\rm{snow}})}{v_{\rm{gas}}}
\end{equation}
where $(r_{\rm{peak}} - r_{\rm{snow}}) / v_{\rm{gas}}$ is the typical timescale on which the peak forms, as discussed in \se{timedependent}. The value of $M_{\rm{solids,needed}}$ is similar to the amount of solids that gets `locked up' in the peak\footnote{We stress again that the model assumes that the incoming pebble mass flux is constant over time (see \se{timedependent}). However, the amount of solids needed to form the peak is independent of this assumption.}. \Fg{contour-plots}a shows that the higher the value of $\mathcal{F}_{\rm{s/g}}$, the easier streaming instability can be triggered, as one may expect. It also shows that in order to reach midplane solids-to-gas ratios of $\sim$$10\%$, a solids-to-gas accretion rate $\mathcal{F}_{\rm{s/g}}$ of the same order is required. This also holds for other values of the gas accretion rate $\dot{M}_{\rm{gas}}$ and initial pebble size $\tau_3$. 

\Fg{contour-plots}b is the same as \fg{nml-contour-plots}b, except that the black contours now denote constant values of the peak midplane solids-to-gas ratio. Just like in \fg{contour-plots}a, orange contours denote constant $M_{\rm{solids,needed}}$, but now with $\dot{M}_\mathrm{gas}$ as the parameter on the $x$-axis. The region with closely-spaced contours in \fg{contour-plots}a and the oblique regions in the otherwise fairly horizontal contours in \fg{contour-plots}b are again due to the transition between the Epstein and Stokes drag regimes.

Up to this point we have kept the parameter $\tau_{3}$ constant at its fiducial value of 0.03. Recently, \citet{2016arXiv161107014Y} have shown that the streaming instability mechanism can also work for small particles ($\tau \sim 10^{-3}$). In \fg{taudependence} we now vary $\tau_{3}$ and plot the resulting peak midplane solids-to-gas ratio, whilst fixing $\dot{M}_{\rm{gas}}$ at $10^{-8} M_{\odot} \: \rm{yr}^{-1}$, $\mathcal{F}_{s/g}$ at 0.4, and $\alpha$ at $3 \times 10^{-3}$. Different regimes in the plot are denoted by Roman numbers I-IV. In region I, the particles are well-coupled to the gas. The solids-to-gas ratio flattens out towards low $\tau_{3}$-values, because radial drift and settling do not play a role for these particles. The boundary between regime I and regime II corresponds to the value of $\alpha$. The highest solids-to-gas ratios are reached for particles with stopping times of the same order as $\alpha$ or smaller. For higher values of $\tau$, the larger drift velocity (proportional to $\tau$) starts to dominate over the larger degree of settling (proportional to $\sqrt{\tau}$). Therefore, in regime II, the solids-to-gas ratio decreases with increasing $\tau$. The transition from regime I and regime II corresponds to the $\tau_{3}$-value for which $\epsilon$ is unity. The $\epsilon$ parameter (introduced in \se{summarysam}) increases with increasing $\tau_{3}$, and the peak surface density decreases at a faster rate with increasing $\epsilon$ for $\epsilon > 1$ than for $\epsilon < 1$ (\eq{sigpeak}). The physical explanation is that the solution changes from diffusion-dominated ($\epsilon < 1$) to drift-dominated ($\epsilon > 1$).
Therefore, the solids-to-gas ratio decreases more steeply in regime III than in regime II. The final transition, from regime III to regime IV, is caused by the transition from the Epstein regime to the Stokes regime. For $\tau_{3}$ values of $\sim$$1$, the enhancement effect becomes minimal and the peak solids-to-gas ratio flattens out.

\begin{figure}[t]
	\centering
		\includegraphics[width=0.495\textwidth]{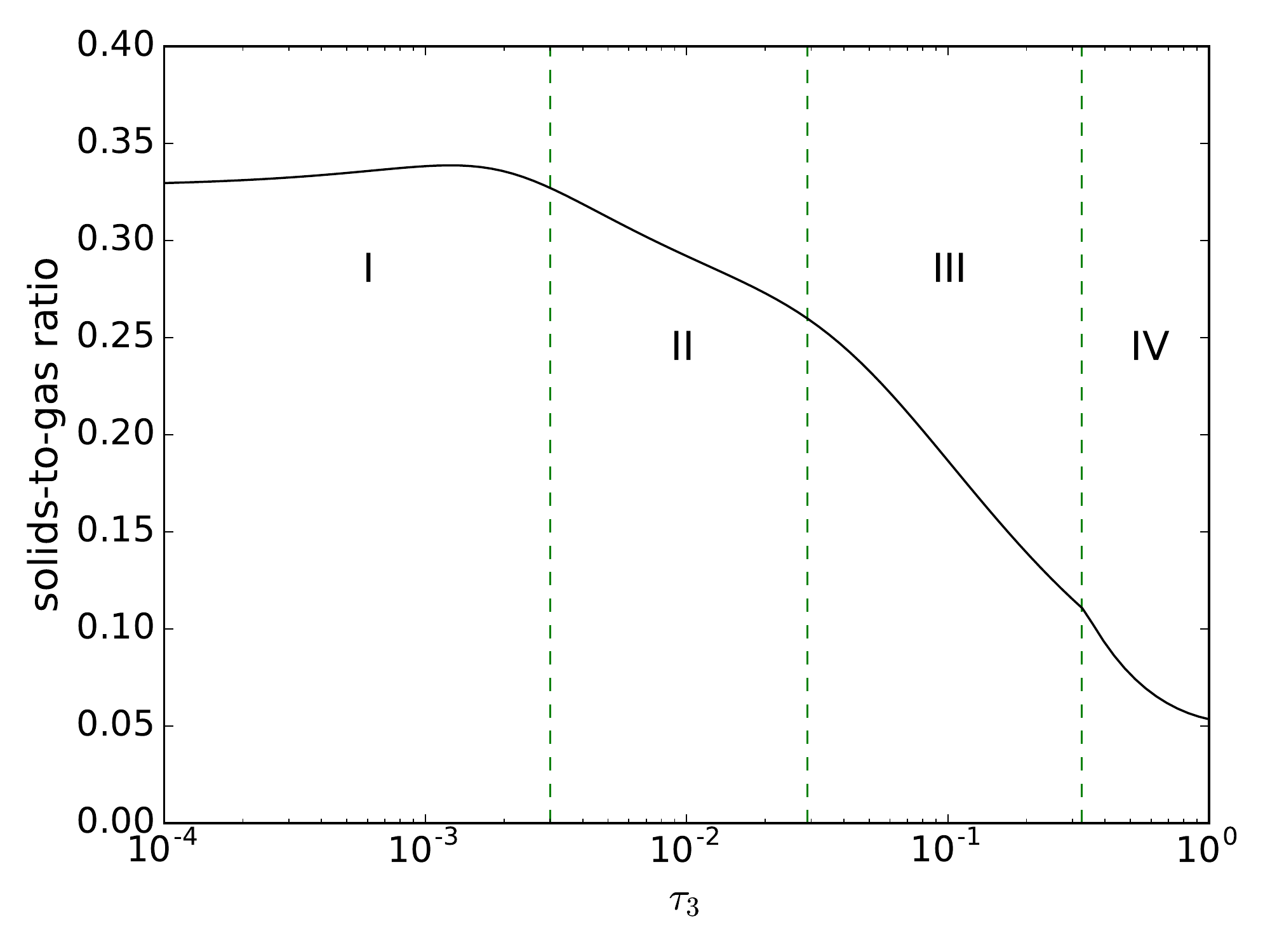}
        \caption{Peak midplane solids-to-gas ratio as function of $\tau_{3}$. $\dot{M}_{\rm{gas}}$ is fixed at $10^{-8} M_{\odot} \: \rm{yr}^{-1}$, $\mathcal{F}_{s/g}$ at 0.4, and the value of $\alpha$ is $3 \times 10^{-3}$. We can distinguish four regimes I-IV, which are explained in the main text.\label{fig:taudependence}}
\end{figure}

From \fg{contour-plots}a and \fg{contour-plots}b we learn that in the context of the $\alpha$-viscosity disk model, the largest solids-to-gas ratios near the snowline are reached for high $\alpha$-values. This means that our diffusion model does not require a very quiescent disk to achieve high solids-to-gas ratio, and also works in the early disk phase when $\alpha$ and $\dot{M}_{\rm{gas}}$ are presumably high. Our results show that for high $\alpha$ and high $\dot{M}_{\rm{gas}}$, tens of Earth masses are locked up in pebbles in an annulus outside the snowline with a width of the order of 1 au.

\section{Discussion}\label{sec:discussion}
\subsection{Conditions for streaming instability}
Our results show that the ice surface density can be enhanced by a factor $\sim$$3$--$5$ outside the snowline, due to the effect of water diffusion and subsequent condensation onto icy pebbles (adopting $\zeta < 0.5$ or the many-seeds model increases the enhancement by another factor $\sim$$2$). This differs from the results of \citet{1988Icar...75..146S}, who found that the same effect could lead to an enhancement by as much as 75. However, they considered a static system, whereas our model accounts for radial drift of pebbles and gas accretion onto the star. Even though we found a modest enhancement, we showed that the pebble density at the midplane can reach tens of percent of the gas density, under the condition that the solids-to-gas accretion rate $\mathcal{F}_{\rm{s/g}}$ is of order 0.1. 

But what are realistic values for $\mathcal{F}_\mathrm{s/g}$? \citet{2016A&A...591A..72I} have calculated the pebble mass flux $\dot{M}_{\rm{peb}}$ in a viscously-evolving disk, taking into account dust growth and radial drift. They found that after $10^6$~yr, $\mathcal{F}_{\rm{s/g}} \sim 0.3$ for a turbulence parameter $\alpha = 10^{-3}$ and a gas accretion rate $\dot{M}_{\rm{gas}} = 10^{-8} M_{\odot} \: \rm{yr}^{-1}$. Since this result suggests $\mathcal{F}_{\rm{s/g}}$-values of order 0.1 are plausible, water condensation can plausibly trigger streaming instability near the snowline.
According to \citet{2016A&A...591A..72I}, $\dot{M}_{\rm{peb}}$ is inversely proportional to $\alpha$. This implies that even though our results show that for high $\alpha$-values, water diffusion and condensation leads to the largest solids-to-gas ratio at the snowline, a smaller pebble mass flux is expected, which reduces the effect. The pebble mass flux is, however, dependent on time $t$ as $\dot{M}_{\rm{peb}} \propto t^{-1/3}$, meaning that the pebble mass flux earlier in the disk is higher than the fiducial $30\%$ of the gas accretion rate quoted before.
Because higher $\alpha$-values lead to larger solids-to-gas ratios but to lower pebble mass fluxes, we conclude that intermediate $\alpha$-values ($\sim$$10^{-3}$) are most favorable to trigger streaming instabilities. The fact that the diffusion-driven model of this work already produces enhancement of solids at intermediate $\alpha$-values and high $\dot{M}_\mathrm{gas}$, implies that planetesimal formation does not have to await the arrival of quiescent disk conditions. Provided a sufficiently large pebble flux, our results imply that planetesimals can form early.

\subsection{Interior or exterior?}
In \citet{2011ApJ...728...20S}, it is argued that the evaporation of icy pebbles leads to a large enhancement of small silicate grains interior to the snowline. In their model, micron-sized silicate grains are instantaneously released at the snowline, and then follow the radial motion of the gas. Since the gas moves at a much lower velocity than the drift velocity of the icy pebbles, the silicate grains pile-up and become gravitationally unstable. A similar mechanism for the formation of planetesimals in the inner disk has recently been presented by \citet{2016A&A...596L...3I}.

We confirm that the solids-to-gas ratio interior to the snowline is boosted due to the slow radial motion of the small silicates (see \fg{ss-fid-clean}c), but find smaller enhancements than \citet{2011ApJ...728...20S} and \citet{2016A&A...596L...3I}. These works assumed that the evaporated silicate grains remained in the same vertical layer as the icy pebbles, which had settled to the midplane due to their larger size. This is a natural assumption if one adopts instantaneous evaporation. However, in our model silicate grains are not instantaneously released from the pebbles at a single location, but are instead released across the ice evaporation front -- the region between the snowline and the peak radius. With the exception of very low $\alpha$-values, the width of the ice peak is (much) larger than the gas scale height. Consequently, the timescale for vertical mixing is smaller than the timescale on which the silicate grains traverse the evaporation front, justifying our assumption that silicate grains (and water vapor) are distributed over $H_\mathrm{gas}$. Allowing for the vertical mixing of silicate grains is why we found smaller solids-to-gas ratios interior to the snowline than \citet{2011ApJ...728...20S} and \citet{2016A&A...596L...3I}.

In this work we did not model the porosity of pebbles, but assumed pebbles are compact spheres. 
For homogeneous porous spheres, the results are the same as for compact spheres with the same stopping time, if they are in the Epstein regime. This is because the evaporation and condensation rates (\eq{ecrates}--(14)) depend on $s_{\rm{p}}^2 / m_{\rm{p}}$ (the surface to mass ratio of a pebble), which is effectively the Epstein stopping time (\eq{stoppingtime}). For fractal aggregates, however, the evaporation rate could conceivably be increased due to a larger total surface area available for evaporation. In that case, the location of the snowline ($r_{\rm{snow}}$) and of the peak ($r_{\rm{peak}}$) are pushed to larger radial distances, because drifting pebbles start to evaporate earlier. The shape of the solids distribution profile and hence the width of the evaporation front, however, remain similar. We have confirmed this behaviour by running our simulation with an evaporation rate that has been arbitrarily increased by a factor 100 compared to \eq{ecrates}. Therefore, a fractal interior structure of the pebbles would not much change the extent of the region in which pebbles evaporate. We note however that in the many-seeds model, a large increase in opacity due to small silicate grains might lead to a steeper temperature profile than assumed here, and therefore to a more narrow evaporation front.

Even though we found smaller enhancements interior to the snowline, adopting the many-seeds model leads to larger enhancements outside the snowline compared to the single-seed model. This is because outward radial diffusion and subsequent sticking of silicate grains to icy pebbles adds to the ice peak and therefore enforces the enhancement of the solids-to-gas ratio outside the snowline (see \fg{ss-fid-cleanvscomplete}).

\subsection{Condensation onto small grains}\label{sec:disc-condens}
In our single-seed model, we adopted a single-size assumption for the solid particles, applicable to a drift-limited scenario, in which particles grow until they decouple from the gas and start to drift inward (see, {\it e.g.}, \citet{2010A&A...513A..79B,2016A&A...586A..20K}). Because the solids density is larger close to the star than in the outer disk, particles close to the star start to drift inward earlier than particles in the outer disk. This leads to an inside-out clearing of the solids component of the disk. In this case, one would not expect small particles to be present in significant amounts near the snowline at the time when icy pebbles approach the snowline, drifting in from the outer disk. Therefore, in the drift-limited case the single-size approximation for pebbles is valid.

The single-size assumption does not hold for a fragmentation-limited size distribution, however. In that case, pebbles dominate the total mass of solids, but small particles dominate the total surface area (see, {\it e.g.}, \citet{2012A&A...539A.148B}). Then, condensation of water vapor will predominantly happen onto small particles, instead of onto pebbles as considered in this work. Also, in our many-seeds model, we do not allow for condensation onto the small grains that get released from icy pebbles upon evaporation.

Within the single-seed framework, we can mimic the situation where condensation happens onto small grains rather than onto pebbles by adopting a very small typical stopping time just outside the snowline ($\tau_{3}$). This results in approximately the same or even larger peaks, as illustrated by \fg{taudependence}. In that sense, assuming condensation occurs solely onto pebbles, rather than onto small grains, is a conservative choice. However, a more correct approach would allow for condensation onto small grains in the many-seeds model, in which icy pebbles break up into small grains upon evaporation, or, alternatively, within a fragmentation-limited scenario by adopting a full size distribution. Such improvements will be considered in future work.

\subsection{Relevance of coagulation}
The many-seeds model ignores coagulation among the small silicate grains, as well as the possibility that water vapor condenses on to silicates.
Because there are so many of them, the average silicate grain size would not increase significantly due to water condensation. In that case, the assumption that they are well-coupled to the gas would still hold. However, growth of silicates might also occur due to coagulation. Growth increases the drift velocity, thereby reducing the effect of the outward-diffusion and recondensation of the silicate component.
Consequently, coagulation among silicates would bring the many-seeds model closer to the single-seed model due to a decrease in the enhancement of `locked' silicates outside the snowline. A similar trend is expected when the silicates that are released upon evaporation are larger. 
According to the experimental work of \citet{2011MNRAS.418L...1A}, an icy aggregate of 1 cm in size breaks up into about 200 sub-aggregates when it evaporates. In any case, we expect that a more realistic `break-up' or coagulation model will lead to results that lie in between the results for our many-seeds and single-seed model designs.

In our work, we have also neglected coagulation among pebbles. On the one hand, one could think that icy pebbles on their way to the snowline would grow even more because of coagulation -- at least as long their relative velocities stay low enough. On the other hand, according to \citet{2011ApJ...733L..41S} and \citet{2016ApJ...821...82O}, the fragmentation threshold velocity for icy pebbles near the snowline is reduced due to sintering. This might motivate a smaller average icy pebble size (a lower $\tau_3$ in our nomenclature) and therefore a smaller radial drift velocity, which would further increase the solid surface density exterior to the snowline (see \fg{taudependence}).

\subsection{Observational implications}
We showed that intermediate-to-high $\alpha$-values are most favorable for triggering streaming instability near the snowline. The onset of streaming instability manifests itself through the formation of a peak in the solids-to-gas ratio that grows in height (increasing $\Sigma_\mathrm{ice}$) but also in width -- the latter because collective effects cause pebbles to move slower. Even though more solids are required for high $\alpha$ and high $\dot{M}_{\rm{gas}}$ than for low $\alpha$ and low $\dot{M}_{\rm{gas}}$, tens of Earth masses are sufficient to form the ice peak. This also means that tens of Earth masses can be stored in an annulus outside the snowline, especially when the system is characterized by high pebble and gas accretion rates. Such annuli of solid enhancement can be observable with facilities as ALMA. In the context of the snowline, the ring structure seen in the TW Hya system with a width of about 1 au \citep{2016ApJ...820L..40A} might very well correspond to a realisation of the water-diffusion effect in a highly turbulent disk with a high gas accretion rate.

\bigskip
The fact that the planets in the inner Solar System are water-poor \citep{2012E&PSL.313...56M,2012LPI....43.1121M}, even though the snowline should have migrated to 1 au before the disk was depleted, was pointed out as the `snowline problem' by \citet{2011ApJ...738..141O}. The early formation of a protoplanet near the snowline might provide a solution to this problem, as proposed recently by \citet{2016Icar..267..368M}. Once a protoplanet near the snowline has formed it could halt the inward-drifting pebble flux, either by accreting the pebbles \citep{GuillotEtal2014}, or by trapping them in pressure maxima created by the newborn planet \citep{2014ApJ...785..122Z, 2014A&A...572A..35L}. In the meantime, the water vapor gets accreted to the star and the snowline location migrates inward ({\it e.g.}, \citet{2011ApJ...738..141O}. Consequently, an early formation of a protoplanet near the snowline might explain the lack of water in the inner Solar System. The water diffusion/condensation effect discussed in this paper provides a way to form planetesimals near the snowline in an early stage of the disk (\textit{i.e.}, when the snowline is still located outside of the current water-poor region of the Solar System). After planetesimals of a certain size have formed, they can subsequently accrete pebbles and quickly grow to larger bodies ({\it e.g.}, \citet{OrmelKlahr2010, 2012A&A...544A..32L, 2014A&A...572A.107L, 2015A&A...582A.112B, 2016A&A...586A..66V}). Therefore, our water diffusion/condensation model provides the first key step to realize the early formation of a protoplanet near the snowline. 

\section{Conclusions}\label{sec:conclusions}
Our main findings can be summarized as follows.
\begin{enumerate}
    \item Water diffusion and condensation near the snowline can result in an enhancement of a factor several in the ice surface density. Because of the dynamic setup of our model where water vapor is carried away with the accreting gas, this enhancement is much less than found by \citet{1988Icar...75..146S}. Nevertheless, the boost in the solids-to-gas ratio can still trigger streaming instability, provided a large enough pebble flux.
    \item The peak in the ice surface density is not located at the snowline, but exterior to it ($r_\mathrm{peak}>r_\mathrm{snow}$). With larger incoming pebble flux, the peak becomes broader, because the back-reaction of the solids on the gas reduces the pebble drift velocity. Depending on the turbulence strength, the width of the peak can be as much as $\sim$1 au.
    \item Such broad peaks can contain tens of Earth masses in pebbles, appearing as bright annuli at radio wavelengths.
    \item The release of many micron-sized silicate grains upon evaporation of icy pebbles produces a peak in `locked' silicates exterior to the snowline, due to diffusion and sweep-up of the silicates. This peak adds to the ice peak and therefore enforces the enhancement of the solids-to-gas ratio outside the snowline.
    \item Interior to the snowline, the solids-to-gas ratio is boosted if many micron-sized silicate grains are released during evaporation, because they cause a `traffic jam' effect. However, because silicate grains mix with the background gas before they cross the snowline, the solids-to-gas ratio enhancement just interior to $r_\mathrm{snow}$ is limited.
    \item In the context of a viscous disk model, the ratio between the accretion rate of solids and gas needs to be of order $\sim$$0.1$ in order to reach solids-to-gas midplane ratios of the order of tens percent. The mechanism operates best at intermediate ($\sim$$10^{-3}$) $\alpha$-values. Therefore, planetesimals can form at an early time in the evolution of the disk.
   \end{enumerate}

\begin{acknowledgements}
    D.S.\ and C.W.O\ are supported by the Netherlands Organization for Scientific Research (NWO; VIDI project 639.042.422). We would like to thank Jonathan Guyer and Daniel Wheeler for their elaborate answers to \texttt{FiPy}-related questions. We would also like to thank Carsten Dominik, Shigeru Ida, Satoshi Okuzumi, Anders Johansen, Daniel Carrera and the other participants of the Lorentz workshop `New directions in Planet Formation' (Leiden, July 2016) for stimulating discussions. We are indebted to Sebastiaan Krijt, Jeff Cuzzi, and the anonymous referee for constructive feedback on the manuscript.
\end{acknowledgements}

\bibliography{AApaperbib_v2}

\newpage
\appendix

\section{List of symbols}
\begin{table}
    \caption{List of frequently-used symbols\label{tab:symbols}}
    \begin{tabular}{ll}
        \hline
        \hline
        $\mathcal{M}$       & mass flux \\
        $\mathcal{F}_\mathrm{s/g}$ & solids-to-gas accretion rate \\
        $\Omega$            & Keplerian orbital frequency \\
        $\Sigma$            & surface density \\
        $\alpha$            & turbulent strength parameter \\
        $\mu$               & mean molecular weight \\
        $\nu$               & viscosity \\
        $\rho$              & midplane density \\
        $\rho_\bullet$      & internal density \\
        $\tau_S$            & dimensionless stopping time \\
        $\zeta$             & dust fraction in pebbles \\
        $D$                 & gas or particle diffusivity \\
        $H$                 & gas or pebble scaleheight \\
        $\dot{M}$           & accretion rate \\
        $P$                 & pressure \\
        $R$                 & condensation or evaporation rate \\
        $T$                 & temperature \\
        $m$                 & mass \\
        $m_\mathrm{core}$  & silicate mass in single-seed model \\
        $r$                 & disk orbital radius \\
        $s_{\rm{p}}$               & particle (pebble) radius \\
        $v_\mathrm{peb}$    & pebble drift velocity \\
        \hline
    \end{tabular}
\end{table}
A list of frequently-used symbols is given in \Tb{symbols}.

\section{Expression Collective Effects}
\newcommand{\bs}{\boldsymbol}
\label{app:collective}

Expressions for the aerodynamic drift of single particles have been derived by \citet{Whipple1972} and \citet{Weidenschilling1977} for non-accreting disks. Later, \citet{NakagawaEtal1986} accounted for the backreaction forces that the particles collectively exert on the gas. The gas motion is then outwards while the total dust+gas mass flux becomes zero. These expressions can be generalized to account for a size distribution of particles \citep{TanakaEtal2005,EstradaCuzzi2008,BaiStone2010i}. 

In (steady) accretion disks gas moves inwards at a rate that is set by the viscosity. This modifies the single particle drift expressions. In particular, small particles, which show negligible radial drift, nevertheless move inwards as they are carried by the gas. While for the single particle case the modification of the drift expressions for accreting disks is straightforward, expressions accounting for both collective effects and viscosity have, to the best of our knowledge, not yet been presented. We derive those here.

The equations of motions in a frame co-rotating with the local Keplerian period read:
\begin{subequations}
    \label{eq:F-eoms}
	\begin{align}
	\frac{D\bs{v}}{Dt} = {} &  -\frac{\bs{v}-\bs{u}}{t_\mathrm{stop}} -2\bs{\Omega}_K \times \bs{v} +\bs{F}_\mathrm{Euler-dust} \\
	\frac{D\bs{u}}{Dt} = {}& \frac{\rho_p}{\rho_g}\frac{\bs{v}-\bs{u}}{t_\mathrm{stop}} -2\bs{\Omega}_K \times \bs{u} +\bs{F}_\mathrm{Euler-gas} +\bs{F}_\mathrm{\nabla P} +\bs{F}_\nu
	\end{align}
\end{subequations}
where $\bs{v}$ is the particle velocity and $\bs{u}$ the gas velocity, $D/Dt$ is the material derivative, the first term on the RHS is the drag force, the second term is the Coriolis force, and $\bs{F}_\mathrm{Euler}=-(d\mathbf{\Omega}_K/dt)\times\mathbf{r}$ is the Euler force that arises due to the radial motion. It therefore appears in the radial equations. In addition, the equation of motion for the gas includes a hydrostatic correction due to a radial pressure gradient ($\bs{F}_{\nabla P}$) and viscous forces ($\bs{F}_\nu$). Following convention, we write the pressure gradient in terms of a nondimensional $\eta$:
\begin{equation}
    \bs{F}_{\nabla P} = \frac{1}{\rho_g} \frac{dP}{dr} \equiv 2\eta v_K \Omega_K \bs{e}_r
    \label{eq:P-grad}
\end{equation}
and use the thin disk approximation (consider column densities $\Sigma_\mathrm{gas}$ instead of $\rho$) for the viscous force:
\begin{equation}
    \bs{F}_\nu 
    = \frac{1}{\Sigma_\mathrm{gas}} \nabla \cdot \mathrm{\mathbf{T}}
\end{equation}
where $\mathrm{\mathbf{T}}$ is the viscous stress tensor. In cylindrical coordinates, the only relevant term is:
\begin{equation}
    T_{r\phi} = \nu \Sigma_\mathrm{gas} r \frac{d\Omega_K}{dr},
\end{equation}
with which the viscous force becomes
\begin{equation}
    \bs{F}_\nu = \frac{1}{\Sigma_\mathrm{gas}} \nabla \cdot \mathrm{\mathbf{T}}
    = \frac{1}{\Sigma_\mathrm{gas}} \frac{1}{r^2} \frac{\partial}{\partial r} \left(r^2 \nu \Sigma_\mathrm{gas} r \frac{d\Omega_K}{dr} \right) \bs{e}_\phi
\end{equation}
where $\nu$ is the kinematic viscosity. In the $\alpha$-disk model we have that $\nu \Sigma_\mathrm{gas}$ is constant and therefore
\begin{equation}
    \bs{F}_\nu = -\frac{3\nu \Omega_K}{4 r} \bs{e}_\phi.
    \label{eq:F-nu}
\end{equation}

In order to solve for the drift velocities, we put the accelerations on the LHS of \eq{F-eoms} zero, $D\bs{u}/Dt=D\bs{v}/Dt=0$. This is justified, because a change in the drift velocities occurs on a timescale of $\sim$$r/v_r$, which is always much longer than the stopping time. Inserting \eq{P-grad} and \eq{F-nu} into \eq{F-eoms} gives a system of four equations and four unknowns (the velocities). Its solution is:
\begin{equation}
    \label{eq:coll-eqs-final}
    \left(\begin{array}{c}
	v_r \\
	v_\phi \\
    u_r \\
    u_\phi
	\end{array}\right) 
    = \frac{1}{\tau_s^2 +(1+\xi)^2}
    \left(\begin{array}{r}
        -2\eta\tau_s -\frac{3}{2}\tilde{\nu}(1+\xi) \\
         \frac{3}{4}\tilde\nu \tau_s -\eta (1+\xi) \\
         2\eta\tau_s \xi -\frac{3}{2}\tilde\nu(1+\tau_s^2 +\xi) \\
         -\frac{3}{4}\tilde\nu \tau_s \xi -\eta (1+\tau_s^2 +\xi)
	\end{array}\right) 
    v_K
\end{equation}
where $\tau_s = t_\mathrm{stop}\Omega_K$, $\xi=\rho_d/\rho_g$, and $\tilde\nu=\nu/r^2\Omega_K$. Note that in the $\alpha$-prescription $\tilde\nu\sim\alpha \eta$, that is, for $\alpha\ll 1$ the modification due to viscosity is minor.

\section{Semi-analytical model}\label{sec:toy-model}
\newcommand{\HiiO}{\ensuremath{\textrm{H}_2\textrm{O}}}
\newcommand{\mH}{\ensuremath{m_\textrm{H}}}
\newcommand{\Mdotice}{\ensuremath{\dot{M}_\mathrm{ice}}}

In this section we aim for an analytic prescription of the key characteristics of the steady-state ice and vapor distribution for any given disc model -- not necessarily with the same gas and temperature profiles as assumed in this study. We will derive expressions for the peak of the ice surface density and its location, which enable us to determine whether the conditions for streaming instability beyond the snowline can be reached.

\subsection{Location of the snowline}
The location of the snowline, $r_{\rm{snow}}$, is the outer-most radius where the entire incoming ice flux can thermodynamically exist in the gaseous state. Its location is found by equating the equilibrium (saturation) pressure to the vapor pressure of \HiiO\ corresponding to the incoming icy pebble flux, that is, $P_{\rm{eq}} = P_{Z,\mathrm{a}}$, where $P_{Z,\mathrm{a}}$ is the steady-state vapor pressure; \textit{i.e.}, the vapor pressure obtained when all the \HiiO\ is in the gas phase. Therefore, the snowline $r_\mathrm{snow}$ is given by the radius $r$ where
\begin{equation}
    P_\mathrm{eq,0} \exp \left[ -\frac{T_a}{T(r)} \right] 
    = \frac{k_B T}{\mu} \frac{\Sigma_{Z,\mathrm{a}}}{\sqrt{2\pi}H_\mathrm{gas}}
    = \frac{k_B T(r)}{\mu} \frac{\Mdotice/3\pi\nu(r)}{\sqrt{2\pi}H_\mathrm{gas}(r)}
    \label{eq:rsnow-eq}
\end{equation}
see \eq{clauclap} and \eq{sigZan}. 
Given $\dot{M}_\mathrm{ice}$, $\alpha$ and a disk model, we can therefore readily solve (numerically) for $r_\mathrm{snow}$. 

A caveat of \eq{rsnow-eq} is that the role of transport processes are neglected. These could potentially introduce disequilibrium corrections. However, we find that the times to achieve the equilibrium (saturation) pressure is typically much shorter than the transport timescales (by diffusion or drift).

\subsection{The equilibrium density}
Generalizing the above arguments to locations beyond $r_\mathrm{snow}$, we approximate the vapor pressure with the saturation pressure -- an approximation that is again justified as long as the timescales for evaporation (and condensation) are sufficiently short (in other words: the super-saturation level is small). Similarly, we can define a vapor surface density ($\Sigma_\mathrm{eq}$) corresponding to $P_\mathrm{eq}$.  Re-arranging \eq{rsnow-eq} in terms of $\Sigma$ we obtain this equilibrium surface density:
\begin{eqnarray}
    \nonumber
    \Sigma_\mathrm{eq} 
    &\equiv& \frac{\sqrt{2\pi}H_\mathrm{gas} \mu}{k_B T} P_\mathrm{eq} \\
    &=& \Sigma_\mathrm{snow} \left( \frac{r}{r_\mathrm{snow}} \right)^p \exp\left[ \frac{T_a}{T_\mathrm{snow}} \left(1 -\left( \frac{r}{r_\mathrm{snow}} \right)^q \right) \right]
    \label{eq:sigZ}
\end{eqnarray}
where $\Sigma_\mathrm{snow}$ and $T_\mathrm{snow}$ are the vapor surface density and the temperature at the location of the snowline. In \eq{sigZ} we have assumed power-laws for the temperature, gas surface density, and the gas scaleheight. In our case we have $p=7/4$ and $q=1/2$ (see \se{disk-model}).

Taylor-expanding the (weakly-varying) power-laws involved in the expression of $\Sigma_\mathrm{eq}$ around $r_\mathrm{snow}$ gives:\footnote{Taylor-expansion of the exponent, on the other hand, would be valid only over a very limited range, because $a_\mathrm{eq} \gg 1$.}
\begin{equation}
    \Sigma_\mathrm{eq}(x)
    \approx \Sigma_\mathrm{snow} \left( 1 +px \right) \exp \left[ -\frac{qT_ax}{T_\mathrm{snow}} \right]
    = \Sigma_\mathrm{snow} \left( 1 +px \right) e^{ -a_\mathrm{eq} x}
    \label{eq:sigma-eq-approx}
\end{equation}
where $x=(r-r_\mathrm{snow})/r_\mathrm{snow}$ is the fractional distance beyond the snowline and 
\begin{equation}
    a_\mathrm{eq}
    = \frac{qT_a}{T_\mathrm{snow}}
    \approx 15 \left( \frac{q}{0.5} \right) \left( \frac{T_\mathrm{snow}}{200~\mathrm{K}} \right)^{-1}
\end{equation}
is a dimensionless number that weakly depends on the location of the snowline.

\begin{figure}[t]
	\centering
    \includegraphics[width=88mm]{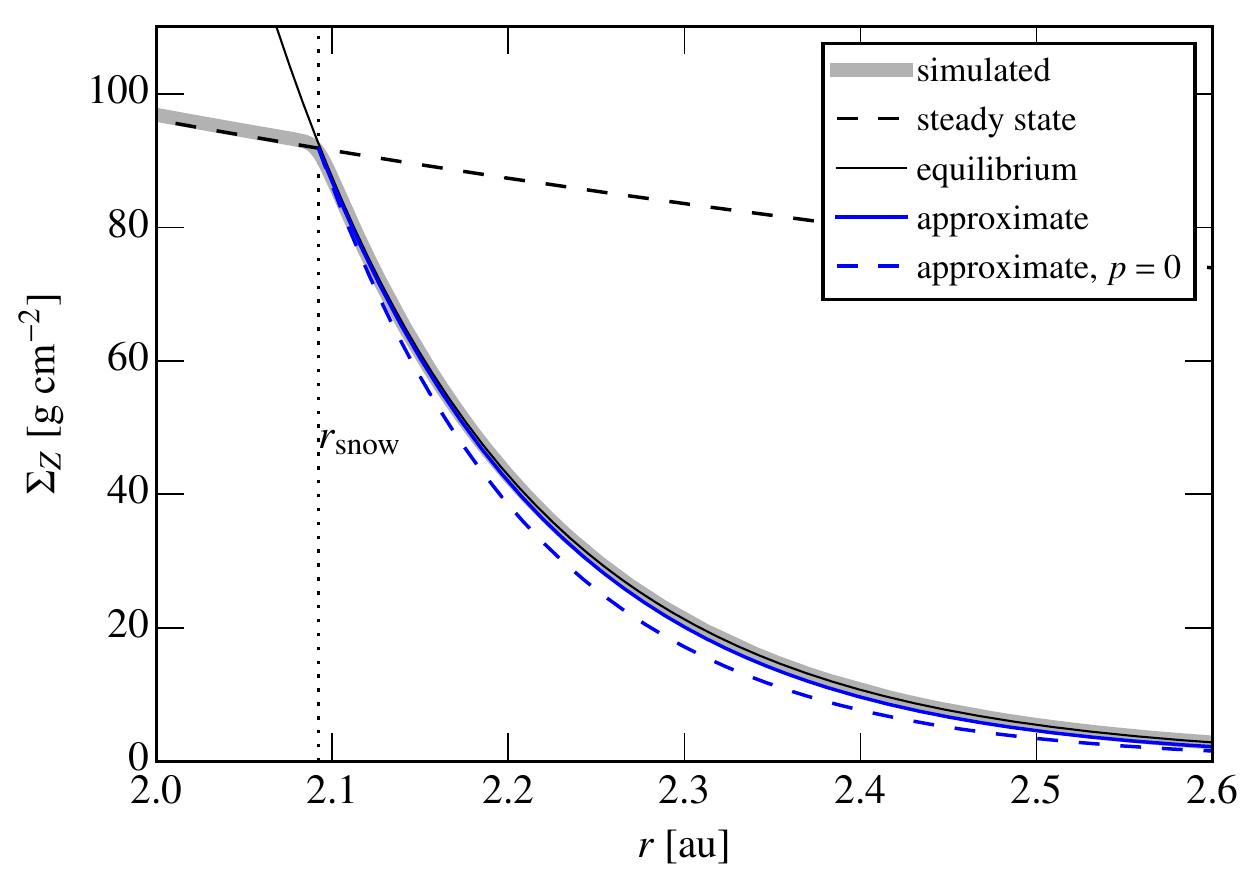}
    \caption{Vapor surface density profiles. Shown in black are the steady state vapor profile when all the ice is in the gas phase (thin solid curve), and the surface density corresponding to equilibrium vapor pressure $P_\mathrm{eq}$ (\eq{sigZ}; dashed solid curve). The snowline $r_\mathrm{snow}$ is located at the intersection of these two curves. The numerically-obtained profile (thick gray curve) closely follows the minimum of steady-state and equilibrium profiles. The blue curves present mathematically-convenient approximations to $\Sigma_\mathrm{eq}$ (see text), valid for $r>r_\mathrm{snow}$.}
    \label{fig:toy1}
\end{figure}
In \fg{toy1} the respective vapor density profiles have been plotted. Parameters correspond to those of the default model (Table~\ref{table:fiducial}). The simulated result of \se{results-fid} is plotted by the thick gray curve. It is characterized by a sharp dent at $r_\mathrm{snow}$. Exterior to the snowline $\Sigma_Z$ follows the equilibrium profile (solid black curve), while interior to $r_\mathrm{snow}$, it is limited by the imposed mass flux (solid dashed curve). The approximate expressions derived in \eq{sigma-eq-approx}, valid for $r>r_\mathrm{snow}$, are shown by the blue curves. We obtain $r_\mathrm{snow}=2.09$ au, $\Sigma_\mathrm{snow}=92~\mathrm{g~cm^{-2}}$ and $a_\mathrm{eq}=16.9$. The  solid blue curve approximates $\Sigma_\mathrm{eq}$ very closely. The dashed blue curves gives a further approximation to $\Sigma_\mathrm{eq}$, obtained by putting $p=0$. This slightly underestimates $\Sigma_Z$ but nevertheless gives a reasonable approximation that we will use below.

Assuming $\Sigma_Z = \Sigma_\mathrm{eq}$ for the vapor density beyond $r_\mathrm{snow}$, we write for the associated mass flux $\mathcal{M}_Z$:\begin{eqnarray}
    \mathcal{M}_Z \approx
    \mathcal{M}_{\mathrm{eq}} 
    \nonumber
    &=& -v_\mathrm{gas}\Sigma_\mathrm{eq} -D_\mathrm{gas} \Sigma_\mathrm{gas} \frac{d}{d r} \left( \frac{\Sigma_\mathrm{eq}}{\Sigma_\mathrm{gas}} \right)  \\
    \label{eq:Mz-eq-def}
    &\approx& -v_\mathrm{gas}\Sigma_\mathrm{eq} -D_\mathrm{gas} \frac{d \Sigma_\mathrm{eq}}{d r}
\end{eqnarray}
where we neglected the curvature of the disk, \textit{i.e.,} we assumed that the scales over which $\Sigma_\mathrm{eq}$ and $\Sigma_\mathrm{gas}$ change are much smaller than $r$. This is justified as long as $x\ll1$. In a similar vein, we assume that quantites as $D_\mathrm{gas}$ and $v_\mathrm{gas}$ do not vary. With \eq{sigma-eq-approx} for $\Sigma_\mathrm{eq}$ we obtain:
\begin{eqnarray}
    \label{eq:Mz-eq}
    \frac{\mathcal{M}_{\mathrm{eq}}}{D_\mathrm{gas}/r_\mathrm{snow}}
    &=& -\frac{v_\mathrm{gas}r_\mathrm{snow}}{D_\mathrm{gas}}\Sigma_\mathrm{eq}- \frac{d\Sigma_\mathrm{eq}}{dx} \\
    \nonumber
    &=& \Sigma_\mathrm{snow}\left( -b_\mathrm{gas} (1+px) + a_\mathrm{eq}(1+px) -p  \right) e^{-a_\mathrm{eq}x}  \\
    &\approx& \Sigma_\mathrm{snow} (a_\mathrm{eq} -b_\mathrm{gas}) e^{-a_\mathrm{eq} x}
\end{eqnarray}
where $b_\mathrm{gas}=v_\mathrm{gas}r_\mathrm{snow}/D_\mathrm{gas}$ is the dimensionless gas velocity and in the last step $p=0$. In a viscously-evolving disk $v_\mathrm{gas}=3\nu/2r$ and hence $b_\mathrm{gas} = 1.5$.

\subsection{The ice profile}
With the equilibrium profile as the approximation to the vapor density, we solve for the surface density of the ice by invoking conservation of mass. In steady-state the total \HiiO\ mass flux $\mathcal{M}_\mathrm{tot,ice} = \mathcal{M}_\mathrm{ice}+\mathcal{M}_Z$ is constant. Interior to the snowline, $\mathcal{M}_Z=\mathcal{M}_\mathrm{tot,ice}$ while for $r\gg r_\mathrm{snow}$ $\mathcal{M}_Z \ll \mathcal{M}_\mathrm{ice}\approx \mathcal{M}_\mathrm{tot,ice}$. But near the snowline the mass fluxes change rapidly. Here, the steep gradient in the vapor density causes an outward (positive) mass flux, which increases the inwardly-directed icy pebble flux:
\begin{equation}
    \label{eq:flux-balance}
    -\mathcal{M}_\mathrm{ice} 
    \equiv \Sigma_\mathrm{ice} v_\mathrm{ice} +D_{\rm{p}}\frac{d \Sigma_\mathrm{ice}}{dr}
    = \mathcal{M}_Z -\mathcal{M}_\mathrm{tot,ice}.
\end{equation}
(Note that just outside $r_\mathrm{snow}$, $\mathcal{M}_\mathrm{ice}$ and $\mathcal{M}_\mathrm{tot,ice}$ are negative and $\mathcal{M}_Z$ is positive.)
With $\mathcal{M}_\mathrm{eq}$ for $\mathcal{M}_Z$ \eq{flux-balance}, in terms of $x$, reads
\begin{equation}
    \frac{d\Sigma_\mathrm{ice}}{dx} +b_\mathrm{peb} \Sigma_\mathrm{ice} 
    = \frac{r_\mathrm{snow}}{D_{\rm{p}}} \left( \mathcal{M}_{\mathrm{eq}} -\mathcal{M}_\mathrm{tot} \right)
    \label{eq:Sigma-ice-eq}
\end{equation}
where $b_\mathrm{peb}=r_\mathrm{snow}v_\mathrm{peb}/D_{\rm{p}}$ is the ratio between the speed at which the pebbles drift in and the radial velocity of the gas. Usually, pebbles outpace the gas and $b_\mathrm{peb}\gg1$.

To solve \eq{Sigma-ice-eq} we will assume that $b_\mathrm{peb}$ and other parameters are constant in $x$. Assuming constant $b_\mathrm{peb}$ is clearly an approximation, since the pebbles may (i) acquire thick icy mantles on their approach to the snowline but (ii) lose all their ice once their are very close to $r_\mathrm{snow}$. Hence, the aerodynamical properties of the pebbles and therefore $v_\mathrm{peb}$ are expected to vary considerably. Disregarding these concerns, momentarily, the solution to \eq{Sigma-ice-eq} reads:
\begin{eqnarray}
    \Sigma_\mathrm{ice}\nonumber
    &=& e^{-b_\mathrm{peb} x} \int_0^x e^{b_\mathrm{peb} x'} \frac{r_\mathrm{snow} \mathcal{M}_{\mathrm{eq}}(x')}{D_{\rm{p}}} dx' \\
    &=& \Sigma_\mathrm{snow}(1-\epsilon_\mathrm{gas}) \left[ \frac{c_\mathcal{M}(1-e^{-a_\mathrm{eq}\varepsilon x})}{\varepsilon} +\frac{e^{-a_\mathrm{eq}x} -e^{-a_\mathrm{eq}\varepsilon x}}{\varepsilon-1} \right] 
    \label{eq:sigice-sol}
\end{eqnarray}
where we used \eq{Mz-eq} for $\mathcal{M}_\mathrm{eq}$, put $p=0$\footnote{For $p\neq 0$ a closed-form solution is possible but extremely contrived.} and defined $\varepsilon=b_\mathrm{peb}/a_\mathrm{eq}$, $\varepsilon_\mathrm{gas}=b_\mathrm{gas}/a_\mathrm{eq}$ and $c_\mathcal{M}=-\mathcal{M}_\mathrm{tot,ice} r_\mathrm{snow} /\Sigma_\mathrm{snow} D_{\rm{p}}(a_\mathrm{eq}-b_\mathrm{gas})$ to keep the notation concise. 
Typically, $\varepsilon_\mathrm{gas}\approx 1.5/a_\mathrm{eq}$ and $c_\mathcal{M}$ are small numbers. In a steady-state viscous disk $\Sigma_\mathrm{snow} \approx \dot{M}_\mathrm{ice}/2\pi r v_\mathrm{gas}$ and $\mathcal{M}_\mathrm{tot,ice} = -\dot{M}_\mathrm{ice}/2\pi r_\mathrm{snow}$. Hence in a viscous disk $c_\mathcal{M}$ is related to $\varepsilon_\mathrm{gas}$:
\begin{equation}
    c_\mathcal{M} 
    = \frac{v_\mathrm{gas}r_\mathrm{snow}}{D_{\rm{p}}(a_\mathrm{eq}-b_\mathrm{gas})} 
    = \frac{\varepsilon_\mathrm{gas}}{1-\varepsilon_\mathrm{gas}}
    \approx{0.1}
\end{equation}

\begin{figure}[t]
	\centering
    \includegraphics[width=88mm]{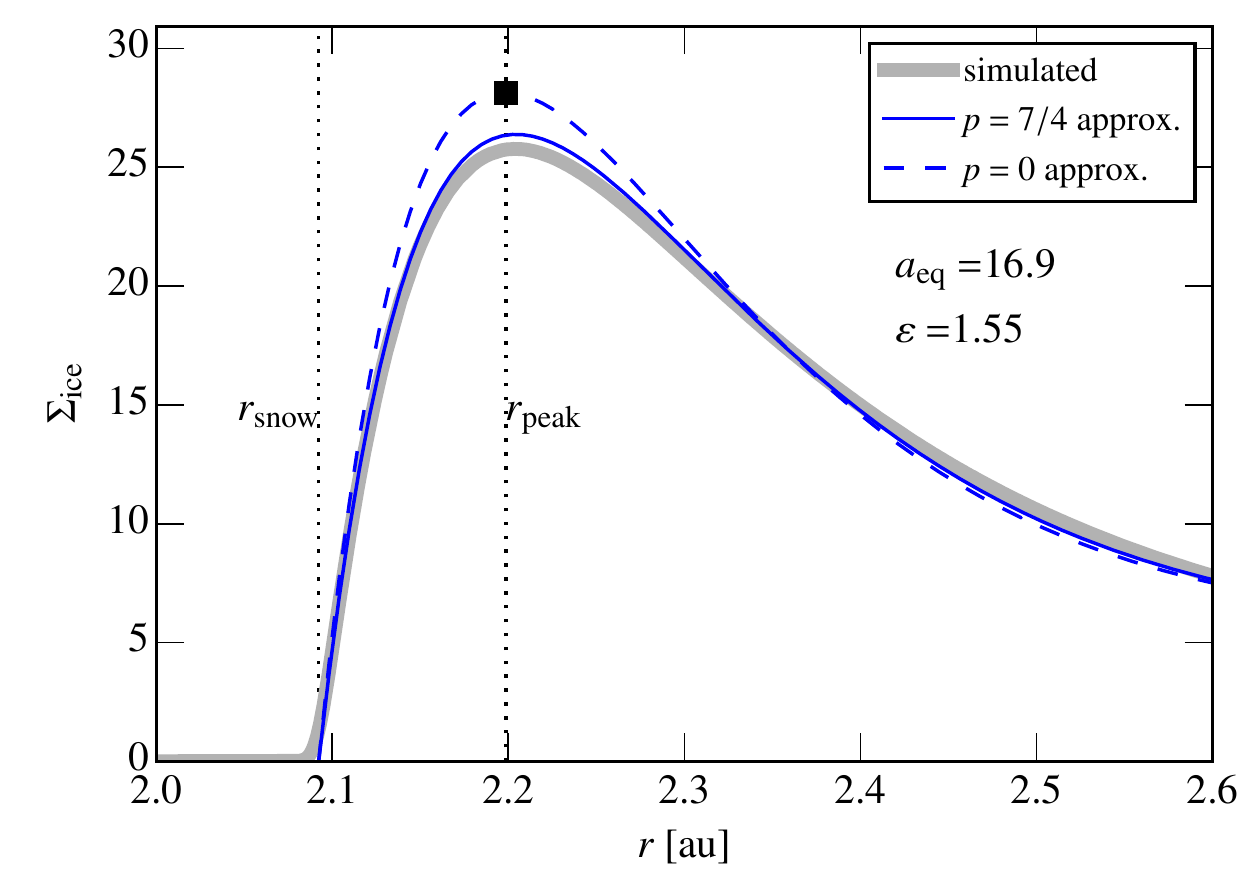}
    \caption{Steady state surface density of ice for the default model parameters. The simulated profile from the numerical model (thick grey line) is well reproduced by the full analytical solution using $p=7/4$ (solid curve; formula not shown in the main text) and the further $p=0$ approximation (dashed curve; \eq{sigice-sol}). The black square denotes the values corresponding to the mass peak ($r_\mathrm{peak}, \Sigma_\mathrm{peak}$) given by \eqs{rpeak}{sigpeak}.}
    \label{fig:toy2}
\end{figure}
In \fg{toy2} we plot $\Sigma_\mathrm{ice}$ for the parameters of the default model (Table~\ref{table:fiducial}).  The simulated data are given by the thick grey curve while the analytical profiles are in blue -- the more precise approximation (with $p=7/4$; solid) and the $p=0$ case in dashed.  The analytical profiles fit the simulation data very well. The key parameter for the analytical profiles is the velocity of the pebbles $v_\mathrm{peb}$.  Here, we have chosen $v_\mathrm{peb}$ to be the pebble velocity at the ice peak $r_\mathrm{peak}$, $v_\mathrm{peb}\approx2.4~\mathrm{m~s}^{-1}$, which provides a decent fit. The fit is somewhat sensitive to the choice of $v_\mathrm{peb}$; for example, adopting $v_\mathrm{peb}(r_\mathrm{peak})$ with the initial value for $m_\mathrm{ice}$ (\textit{i.e.}, without deposition of vapor) would overestimate the numerical curve by 20\%.

\subsection{Peak values and final tuning}
\Eq{sigice-sol} has a maximum at $(x,\Sigma_\mathrm{ice})=(x_\mathrm{peak},\Sigma_\mathrm{peak})$ where
\begin{equation}
    \label{eq:rpeak}
    x_\mathrm{peak} = \frac{\log \left[ \varepsilon +c_\mathcal{M}(\varepsilon-1)\right]}{a_\mathrm{eq}(\varepsilon-1)}
\end{equation}
or $r_\mathrm{peak} = (1+x_\mathrm{peak}) r_\mathrm{snow}$ in dimensional units, and
\begin{equation}
    \label{eq:sigpeak}
    \Sigma_\mathrm{peak} = (1-\varepsilon_\mathrm{gas}) \frac{c_\mathcal{M} +\left[ c_\mathcal{M}(\varepsilon-1) +\varepsilon\right]^{1/(1-\varepsilon)}}{\varepsilon} \Sigma_\mathrm{snow}.
\end{equation}
Since $c_\mathcal{M}$ and $\varepsilon_\mathrm{gas}$ are fixed, the latter expression only depends on $\varepsilon$. Lower $\varepsilon$ -- meaning: a smaller pebble velocity at $r_\mathrm{peak}$ -- increases both the width ($x_\mathrm{peak}$) and the magnitude of the ice peak, resulting in a more pronounced, `fatter' ice bump. 

\begin{table}
    \centering
    \caption{Comparison between analytical and numerical model}
	\begin{tabular}{lllllll}
	\hline
	\hline
	S/F & \multicolumn{2}{l}{parameter} & \multicolumn{2}{c}{Fipy} & \multicolumn{2}{c}{Analytical} \\
	    & \multicolumn{2}{l}{}       & $r_\mathrm{peak}$ & $\Sigma_\mathrm{peak}$ & $r_\mathrm{peak}$ & $\Sigma_\mathrm{peak}$ \\
	\hline
	S & \multicolumn{2}{l}{(default)}  & {$2.20$} & {$26$} & {$2.20$} & {$26$}\\ 
	S & $\alpha$ & {$3\times10^{-4}$} & {$1.77$} & {$12$} & {$1.77$} & {$11$}\\ 
	S & $\alpha$ & {$0.03$} & {$2.75$} & {$5.0$} & {$2.72$} & {$4.7$}\\ 
	S & $\dot{M}$ & {$1\times10^{-7}$} & {$1.80$} & {$110$} & {$1.80$} & {$96$}\\ 
	S & $\tau_3$ & {$0.3$} & {$2.13$} & {$4.2$} & {$2.12$} & {$4.5$}\\ 
	F & \multicolumn{2}{l}{(default)}  & {$2.34$} & {$15$} & {$2.33$} & {$16$}\\ 
	F & $\alpha$ & {$3\times10^{-4}$} & {$1.85$} & {$5.9$} & {$1.86$} & {$5.3$}\\ 
	F & $\alpha$ & {$0.03$} & {$2.95$} & {$2.7$} & {$2.88$} & {$2.5$}\\ 
	F & $\dot{M}$ & {$1\times10^{-7}$} & {$1.90$} & {$66$} & {$1.89$} & {$58$}\\ 
	F & $\tau_3$ & {$0.3$} & {$2.22$} & {$2.0$} & {$2.21$} & {$2.5$}\\ 
	F & $\mathcal{F}_\mathrm{s/g}$ & {$0.6$} & {$2.30$} & {$26$} & {$2.28$} & {$28$}\\ 
	F & $\mathcal{F}_\mathrm{s/g}$ & {$0.8$} & {$2.29$} & {$40$} & {$2.25$} & {$43$}\\ 
	F & $\mathcal{F}_\mathrm{s/g}$ & {$1$} & {$2.30$} & {$60$} & {$2.25$} & {$63$}\\ 
	\hline
	\end{tabular}
    \label{tab:TM}
    \tablefoot{Column entries denote: (S/F) simple (no $\mu$-variations or collective effects) or full model; (parameter) input parameter that is varied and its value; (\texttt{FiPy}) values for the position of the ice peak (in au) and the surface density (in cgs units) for the ice peak; (Analytical) same for the analytical model. }
\end{table}
All that remains is to find an expression for the pebble velocity at the ice peak $v_\mathrm{peb}$, which in turn depends on the amount of ice those pebbles have accreted. 
To obtain $m_\mathrm{peb} = m_\mathrm{core} +m_\mathrm{ice}$ we invoke conservation of the pebble number flux
\begin{equation}
    \label{eq:Np}
    \mathcal{N}_{\rm{p}} 
    = \frac{\dot{M}_\mathrm{sil}/m_\mathrm{core}}{2\pi r v_\mathrm{peb}}
    = \frac{\Sigma_\mathrm{ice}}{m_\mathrm{ice}}.
\end{equation}
Hence, we can obtain $m_\mathrm{ice}$ from $\Sigma_\mathrm{peak}$, calculate the the aerodynamical properties of the pebbles at the peak (\textit{i.e.}, their stopping time) and obtain the pebble velocity $v_\mathrm{peb}$.

With \eq{Np} we have a closed system of expressions to obtain the relevant quantities at the ice peak.  These can best be computed by an iterative scheme, \textit{e.g.},
\begin{enumerate}
    \item From (an initial guess for) $\varepsilon$ obtain $\Sigma_\mathrm{peak}$ and the position of the ice peak $r_\mathrm{peak}$ from the ice flux conservation model, \eq{sigpeak}.
    \item From $\Sigma_\mathrm{peak}$, obtain the ice mass $m_\mathrm{ice}$ at the ice peak from the pebble conservation law, \eq{Np}.
    \item From $m_\mathrm{ice}$ and the gas properties at $r=r_\mathrm{peak}$, compute the stopping time of the pebbles at $r_\mathrm{peak}$. Update the pebble velocity $v_\mathrm{peb}$ and its normalized variant $\varepsilon$.
\end{enumerate}
This scheme can easily be extended with collective and mean molecular weight effects (\textit{i.e.}, the complete model). Finally, in order to obtain a better match to our numerical results we adopt two ad-hoc empirical `fixes'. First we reduce \eq{sigpeak} by 10\%, which accounts for the fact that the $p=0$ approximation tends to overestimate the peak. Secondly, we slightly increase $\varepsilon$ when it drops below unity. This fix approximately accounts for curvature effects (not included in the model) that become apparent when the ice peak becomes broad (at low $\varepsilon$). Hence we take $\varepsilon = \max(\varepsilon\ast, \varepsilon\ast^{0.8})$ where $\epsilon\ast=(v_\mathrm{peb,peak}r_\mathrm{peak}/a_2 D_\mathrm{peak})$ is the uncorrected value. 

In \tb{TM} a comparison of the model with the numerically-obtained peak parameters is given for a number of model parameters. Generally, the agreement is very good; the error in $\Sigma_\mathrm{peak}$ is at most 20\%. Given its crudeness, the analytical model, however, does not always produce a perfect match -- especially not when it comes to the profile. For example the low turbulence runs ($\alpha=3\times10^{-4}$), which imply high gas densities when $\dot{M}_\mathrm{gas}$ is kept the same, still show a good match to $\Sigma_\mathrm{peak}$, but the profile corresponding to \eq{sigice-sol} is very different from the numerical model. The reason for this is that pebbles enter the Stokes drag regime when approaching the ice peak, causing a sharp increase in the stopping time (and $\varepsilon$). The stopping time at the ice peak is in that case not representative for other radii.

\end{document}